\documentclass[aps,preprint,superscriptaddress,nofootinbib]{revtex4-1}
\pdfoutput=1
\usepackage{physics2,derivative,amsmath,amssymb,hyperref,color,graphicx}
\newcommand{\order}{\mathcal{O}}
\newcommand{\tr}{{\rm tr}}
\usephysicsmodule{ab,ab.braket}
\begin{document}
\count\footins = 1000
\title{Regularization of Functional Determinants of Radial Operators via Heat Kernel Coefficients}

\numberwithin{equation}{section}

\allowdisplaybreaks[1]

\author{Yutaro Shoji}
%\footnote{\href{mailto:}{Yutaro.Shoji@ijs.si}}
\affiliation{Jo\v{z}ef Stefan Institute, Jamova 39, 1000 Ljubljana, Slovenia}
\author{Masahide Yamaguchi}
%\footnote{\href{mailto:}{gucci@ibs.re.kr}}
\affiliation{Cosmology, Gravity, and Astroparticle Physics Group, Center for Theoretical Physics of the Universe, Institute for Basic Science (IBS), Daejeon, 34126, Korea}
\affiliation{Department of Physics, Institute of Science Tokyo, Tokyo 152-8551, Japan}
\affiliation{Department of Physics \& Institute of Physics and Applied Physics (IPAP), Yonsei University, Seoul, 03722, Korea}

\preprint{}

\begin{abstract}
    We propose an efficient regularization method for functional determinants of radial operators using heat kernel coefficients.
    Our key finding is a systematic way to identify heat kernel coefficients in the angular momentum space.
    We explicitly obtain the formulas up to sixth order in the heat kernel expansion, which suffice to regularize up to 13-dimensional functional determinants. We find that the heat kernel coefficients accurately approximate the large angular momentum dependence of functional determinants, and make numerical computations more efficient.
    In the limit of a large angular momentum, our formulas reduce to the Wentzel-Kramers-Brillouin formulas in previous studies, but are extended to higher orders. All the results are available in both the zeta function regularization and the dimensional regularization.
\end{abstract}

\maketitle

%%%%%%%%%%%%%%%%%%%%%%%%%%%%%%%%%%%%%%%%%%%%%%%%%%%%%%%%%%%%%%%%%%%%%%
\section{Introduction}
In quantum field theory, leading order quantum corrections appear in the form of functional determinants, and a lot of effort has been devoted to their evaluation in theoretical and phenomenological studies. A bubble nucleation rate, or a vacuum decay rate, is among the examples where functional determinants play an important role, as they determine the overall factor.
However, the calculation of such functional determinants is not easy because they are calculated in a nontrivial background of a decaying scalar field.
Fortunately, we can take a spherically symmetric background as such a configuration gives the smallest Euclidean action \cite{Coleman:1977th,Blum:2016ipp,Oshita:2023pwr}. It allows us to decompose fluctuations into partial waves, and a functional determinant is expressed as an infinite number of subdeterminants of radial operators. Even with this simplification, the calculation is much more difficult than in the trivial background.

One of the difficulties comes from the divergent ultraviolet (UV) behavior of functional determinants. While the contributions from each subdeterminant are finite, a divergence appears when all of the subdeterminants are multiplied. To get a finite value, we must regulate the determinant using a well-defined scheme such as the dimensional regularization and the zeta function regularization.
It is however not easy to apply such a regularization directly because these schemes utilize an analytic function of either the spacetime dimensions or the argument of the zeta function.
On the other hand, subdeterminants are usually calculated numerically up to a finite angular momentum, and their analytic expressions are not accessible.
To overcome this problem, the {\it subtract and add-back} technique has been adopted in the literature, where one computes a finite quantity by subtracting a reference series having the same UV behavior, and then adds back the regularized value of the reference series.

One such reference series can be obtained using the so-called Feynman diagrammatic approach, which is based on the expansion of the one-loop diagram with mass insertions \cite{Baacke:2003uw}. Since inserting interaction vertices introduces more propagators, only the first few terms are divergent. These terms are then expressed in the angular momentum space, and are used as a reference series. Meanwhile, the add-back terms can be computed in the conventional Feynman diagrammatic way with the dimensional regularization. The diagrams with one and two insertions are relatively easy to calculate, but those with three or more insertions are significantly more difficult. Fortunately, only up to two insertions suffice to regularize the determinant up to five spacetime dimensions, and this approach has been used in various phenomenological applications in four dimensions \cite{Isidori:2001bm,Branchina:2014rva,Endo:2015ixx,Oda:2019njo,Andreassen:2017rzq,Chigusa:2017dux,Chigusa:2018uuj,Khoury:2021zao,Chigusa:2022xpq,Chigusa:2023mqy,Chauhan:2023pur,Baratella:2024hju}.

Another approach is to use a large angular momentum expansion together with the zeta function regularization \cite{Dunne:2004sx,Dunne:2005te,Dunne:2005rt,Dunne:2006ct} (see also a lecture note by Dunne \cite{Dunne:2007rt}). Since it is derived from the Wentzel-Kramers-Brillouin (WKB) approximation, it is also referred to as the WKB approach\footnote{In this work, we avoid the term ``WKB approach'' since the reference series is ultimately obtained via a large angular momentum expansion, regardless of the derivation method.}.
In \cite{Dunne:2006ct}, the reference series is identified in two, three and four dimensions. The virtue of this approach is that the formulas have a remarkably simple structure; they are expressed as simple one-dimensional integrals such as $\int\odif{r}r^{2n-1}[(m^2)^n-(\hat m^2)^n]$, where $m^2$ and $\hat m^2$ are second derivatives of the potential. This not only helps in the analytical evaluation of functional determinants, but also reduces the effort of writing numerical code, which has attracted attention in several applications \cite{Branchina:2013jra,Ai:2023yce,Ekstedt:2023sqc,Ivanov:2022osf,Benevedes:2024tdq,Matteini:2024xvg,Baratella:2025dum}. However, it is challenging to identify such reference series in six and higher dimensions as it is difficult to identify higher-order terms of the large angular momentum expansion.

A similar reference series has been proposed in \cite{Hur:2008yg}, where the results are given in the summed form. While \cite{Dunne:2006ct} is essentially a large angular momentum expansion of the diagrams with mass insertions, \cite{Hur:2008yg} expands the heat kernel of the radial operator at a large angular momentum\footnote{This is different from the heat kernel expansion of the original multi-dimensional operator expressed in the angular momentum space, which we are going to compute in this paper.}. Hence we call it the radial WKB approach.
They obtained explicit formulas that can be used up to five dimensions, but their results imply the formulas will be complicated in higher dimensions.

In this paper, we provide two main results. One is the extension of the Dunne and Kirsten's results \cite{Dunne:2006ct} to higher orders with the help of the heat kernel expansion. 
We provide a systematic way to identify the expression of the trace of each heat kernel coefficient in the angular momentum space. By expanding them at a large angular momentum, we obtain the reference series of \cite{Dunne:2006ct}. We explicitly compute the first six coefficients, which suffice to regularize up to 13 dimensional functional determinants. We found that the expressions are remarkably compact for the first five coefficients. Higher-order results are also important for calculations in lower dimensions, as they accurately approximate large angular momentum contributions and reduce the number of subdeterminants that need to be calculated numerically.

The other main result is the introduction of a new reference series, which offers a remarkably simple and efficient framework for regularizing functional determinants.
We propose to directly use heat kernel coefficients as a reference series. Since they are well-defined beyond the large angular momentum expansion, such a series is expected to behave better than that in \cite{Dunne:2006ct}. In our numerical study, we find that it indeed reproduces the large angular momentum contributions better and reduces the computational cost. We also demonstrate the regularization of up to 13-dimensional functional determinants using the Fubini-Lipatov instantons \cite{Fubini:1976jm}.
 
The above results are obtained in the zeta function regularization, and need to be converted into those in the dimensional regularization, in particular, for phenomenological applications since most coupling constants are available in the $\overline{\rm MS}$ scheme.
It has been known to some extent \cite{Bytsenko:2003tu} that the zeta function regularization gives results similar to dimensional regularization.
However, it is often a source of confusion due to the absence of the pole and the renormalization scale in the zeta function regularization.
Furthermore, the results obtained using the dimensional regularization \cite{Baacke:2003uw} and the zeta function regularization \cite{Dunne:2006ct} look very different, and it is only recently that these expressions have been proven to be analytically identical in four dimensions \cite{Baratella:2025dum}.
We have therefore decided to provide a derivation of the conversion formula for completeness. Our discussion applies to general dimensions and to a general background field, which is not necessarily spherically symmetric.

This paper is organized as follows. In the next section, we introduce functional determinants of radial operators. Then, we review the zeta function regularization and the heat kernel expansion in Sections \ref{sec:zeta} and \ref{sec:hke}. We derive the higher-order terms of the large angular momentum expansion in Section \ref{sec:large_nu}, and introduce our reference series in Section \ref{sec:full_hke}. We apply our method to vacuum decay rates and compare it with the other methods in Section \ref{sec:vacuum_decay}. Finally, we summarize in Section \ref{sec:summary}.

%%%%%%%%%%%%%%%%%%%%%%%%%%%%%%%%%%%%%%%%%%%%%%%%%%%%%%%%%%%%%%%%%%%%%% 
\section{Functional determinant of radial operator}\label{sec:functional_determinant}
We consider a couple of fluctuation operators in $D$-dimensional flat Euclidean spacetime without boundary,
\begin{equation}
    \mathcal M=-\partial^2+m^2(r),~\widehat{\mathcal M}=-\partial^2+\hat m^2,
\end{equation}
where $\hat m^2$ is a non-negative constant, $\partial^2=\partial_\mu\partial^\mu$, $r=\sqrt{x^\mu x_\mu}$ and $\mu=1,\cdots,D$.
The difference of the two operators is denoted by $\delta m^2(r)=m^2(r)-\hat m^2$,
which approaches zero sufficiently fast as $r\to\infty$.
In the following sections, we ignore zero\footnote{The Laplace operator $(-\partial^2)$ does not have a zero mode, as a constant function is not a mode function due to our definition of functional determinants. See the next section for details.} and negative modes since we are interested in the UV behavior of the ratio of determinants. This does not cause any issue because the number of negative modes is typically finite, and we can exclude them from the determinant without changing the UV behavior.
Another way is to use a larger $\hat m^2$ first such that both $\mathcal M$ and $\widehat{\mathcal M}$ do not have zero or negative modes, determine the $\hat m^2$ dependence of the UV divergence, and bring $\hat m^2$ back to the original value. 

We are interested in the ratio of their determinants,
\begin{equation}
    R=\frac{\det\mathcal M}{\det\widehat{\mathcal M}}.
\end{equation}
As is known, this is finite for $D=1$ due to the cancellation of UV contributions between the denominator and the numerator. Since our focus is on the regularization of $R$, we consider $D\geq2$ throughout this paper.

Since the fluctuation operator is $O(D)$ symmetric, it is separable to radial operators using the $D$-dimensional hyperspherical harmonics $Y_{l\chi}(\hat x)$, which satisfies
\begin{equation}
    -r^2\partial^2 Y_{l\chi}(\hat x)=l(l+D-2)Y_{l\chi}(\hat x).
\end{equation}
Here $\hat x=x/|x|$ and $l=0,1,\cdots$ labels the angular momentum.
It is more convenient to use a shifted angular momentum,
\begin{equation}
    \nu=l+\frac{D}{2}-1,
\end{equation}
as this combination appears repeatedly.
Finally, $\chi=1,\cdots,d_\nu$ labels the eigenfunctions having the same $\nu$, and the degeneracy factor is given by
\begin{equation}
    d_\nu=\frac{2\nu\Gamma\ab(\nu-1+\frac{D}{2})}{\Gamma(\nu+2-\frac{D}{2})\Gamma(D-1)}.
\end{equation}
Notice that $d_\nu=2$ for all $\nu$ in two dimensions, which agrees with the limit of $D\to2$ in the above equation.
We will use $\prod_\nu$ and $\sum_\nu$ to indicate a product and a sum over $\nu$ from $\nu=D/2-1$ to infinity.

Then, $R$ is decomposed into subdeterminants as
\begin{equation}
    R=\prod_{\nu} R_\nu^{d_\nu}=\prod_{\nu}\ab(\frac{\det\mathcal M_\nu}{\det\widehat{\mathcal M}_\nu})^{d_\nu},
\end{equation}
where the radial fluctuation operators are defined as
\begin{align}
    \mathcal M_\nu&=-\partial_r^2-\frac{D-1}{r}\partial_r+\frac{\nu^2-\ab(\frac{D}{2}-1)^2}{r^2}+m^2(r),\\
    \widehat{\mathcal M}_\nu&=-\partial_r^2-\frac{D-1}{r}\partial_r+\frac{\nu^2-\ab(\frac{D}{2}-1)^2}{r^2}+\hat m^2.
\end{align}

\section{Zeta function regularization}\label{sec:zeta}
As we have mentioned, the ratio $R$ is divergent for $D\geq2$, and must be regularized to obtain a well-defined value. There are two widely used schemes: one is the dimensional regularization and the other is the zeta function regularization.
We discuss the zeta function regularization first and the dimensional regularization in Section \ref{sec:dimreg}.

We first introduce the zeta function of operator $\mathcal M$, which is defined by
\begin{equation}
    \zeta_{\mathcal M}(s)=\sum_i\frac{1}{\omega^s_i},\label{eq:first_zeta_def}
\end{equation}
where $\omega_i$'s are the eigenvalues of $\mathcal M$, and $i$ runs over all possible eigenvalues.
Strictly speaking, this quantity is divergent independently of $s$ due to infinite spacetime volume. This is obvious, for example, in the case of $\zeta_{-\partial^2}(s)=\int\odif[order=D]{x}\int\frac{\odif[order=D]{k}}{(2\pi)^D}k^{-2s}$.
To make our discussion more precise, we put an artificial boundary at $|x|=r_{\infty}$, and impose Dirichlet boundary conditions for mode functions. Here, the Dirichlet boundary conditions mean that the value of a mode function vanishes at the boundary. Then, we will take $r_\infty\to\infty$ only for quantities that are well-defined in this limit.
In a finite volume, the sum of Eq.~\eqref{eq:first_zeta_def} is convergent for $\Re(s)>D/2$ because $\omega_i$ behaves as $k^2$ at a large momentum $k$.

The difference of two zeta functions, $\zeta_{\mathcal M}(s)-\zeta_{\widehat{\mathcal M}}(s)$ has better convergence because of a cancellation of leading UV contributions, and converges for $\Re(s)>D/2-1$. In addition, if $\delta m^2(r)$ decays sufficiently fast, it effectively serves as a cutoff of the spacetime volume and the difference is well-defined in the limit of $r_\infty\to\infty$.

The zeta function regularization is defined through analytic continuation of the zeta function. Since the first derivative of the zeta function at $s=0$ gives the log of the determinant, the regularized value is given by
\begin{equation}
    \left.\ln \frac{\det\mathcal M}{\det\widehat{\mathcal M}}\right|_{\zeta}=-\lim_{s\to0}\ab[\zeta'_{\mathcal M}(s)-\zeta'_{\widehat{\mathcal M}}(s)].\label{eq:lndet_from_zeta}
\end{equation}
Note that it is regular at $s=0$ unlike the dimensional regularization, and the limit $s\to0$ is well-defined regardless of spacetime dimensions.

For spherically symmetric fluctuation operators, the zeta function is separable as
\begin{equation}
    \zeta_{\mathcal M}(s)=\sum_\nu d_\nu\sum_i\frac{1}{\omega^s_{\nu i}},
\end{equation}
where $\omega_{\nu i}$'s are the eigenvalues of $\mathcal M_\nu$.
Notice that there is no problem with reshuffling the summand since the sum converges absolutely for $\Re(s)>D/2$.

\begin{figure}[t]
    \centering
    \includegraphics[width=0.5\textwidth]{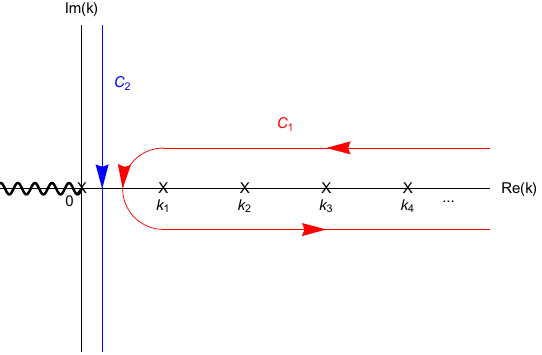}
    \caption{Contours for the integral and the analytic structure of the integrand. The poles of $\odv{}{k}\ln u_\nu(k)$ are at $k=k_i=\sqrt{\omega_{\nu i}}$. The wavy line indicates the branch cut due to $k^{-2s}$.}
    \label{fig:contour}
\end{figure}

A more convenient expression is given as a contour integral \cite{Kirsten:2003py,Kirsten:2004qv},
\begin{equation}
    \zeta_{\mathcal M}(s)=\sum_\nu d_\nu\frac{1}{2\pi i}\int_{C_1}\odif{k}\frac{1}{k^{2s}}\odv{}{k}\ln u_\nu(k),
\end{equation}
where the contour $C_1$ is given in Fig.~\ref{fig:contour}. Here, $\odv{}{k}\ln u_\nu(k)$ is designed to have poles at $k=\sqrt{\omega_{\nu i}}$. Such $u_\nu(k)$ can be constructed in the following way. We first consider a solution of a differential equation,
\begin{equation}
    \mathcal M_\nu f_\nu(\omega;r)=\omega f_\nu(\omega;r),
\end{equation}
with boundary conditions,
\begin{equation}
    \lim_{r\to0}\frac{f_\nu(\omega;r)}{r^{\nu-\frac{D}{2}+1}}=1.
\end{equation}
Then, a function,
\begin{equation}
    u_\nu(k)=f_\nu(k^2;r_{\infty}),
\end{equation}
has zero points at $k^2=\omega_{\nu i}$ because, if it is not zero, the $f_\nu(k^2;r_{\infty})$ does not satisfy the Dirichlet boundary condition, $f_\nu(\omega;r_\infty)=0$. This means that $\odv{}{k}\ln u_\nu(k)$ has poles at $k^2=\omega_{\nu i}$. Although we do not go into details, it is known that $\odv{}{k}\ln u_\nu(k)$ does not have other poles in the $\Re(k)>0$ region.

Finally, we take the difference of two zeta functions,
\begin{equation}
    \zeta_{\mathcal M}(s)-\zeta_{\widehat{\mathcal M}}(s)=\sum_\nu d_\nu\frac{1}{2\pi i}\int_{C_1}\odif{k}\frac{1}{k^{2s}}\odv{}{k}\ln \frac{u_\nu(k)}{\hat u_\nu(k)},\label{eq:zeta_robust}
\end{equation}
with $\hat u_\nu(k)$ being $u_\nu(k)$ calculated with $\widehat{\mathcal M}$. We take this expression as a more robust definition of zeta functions and functional determinants in our paper. We note that the above discussion can be generalized to the cases where there are non-positive modes \cite{Kirsten:2004qv}.

It is, however, often difficult to directly compute Eq.~\eqref{eq:zeta_robust} to obtain a fully analytic function of $s$.
To deal with this problem, the subtract and add-back technique has been adopted in the literature.
We introduce a reference series, $\eta_\nu(s)$, and rewrite Eq.~\eqref{eq:zeta_robust} as
\begin{equation}
    \zeta_{\mathcal M}(s)-\zeta_{\widehat{\mathcal M}}(s)=\sum_\nu\ab[d_\nu\frac{1}{2\pi i}\int_{C_1}\odif{k}\frac{1}{k^{2s}}\odv{}{k}\ln \frac{u_\nu(k)}{\hat u_\nu(k)}-\eta_\nu(s)]+S(s).
\end{equation}
Here, $\eta_\nu(s)$ is chosen such that the terms in the parentheses become finite around $s=0$, and the add-back term $S(s)=\sum_\nu\eta_\nu(s)$ is known analytically. Now, we can bring $s$ around $s=0$ and the contour can be deformed into $C_2$ in Fig.~\ref{fig:contour} for $-1/2<\Re(s)<1/2$ \cite{Kirsten:2003py,Kirsten:2004qv}. Breaking the integral into two parts and moving $C_2$ closer to the imaginary axis, one finds
\begin{align}
    \frac{1}{2\pi i}\int_{C_1}\odif{k}\frac{1}{k^{2s}}\odv{}{k}\ln \frac{u_\nu(k)}{\hat u_\nu(k)}&=\frac{\sin(\pi s)}{\pi}\int_0^\infty\odif{k}\frac{1}{k^{2s}}\odv{}{k}\ln \frac{u_\nu(ik)}{\hat u_\nu(ik)}\nonumber\\
    &=-s\ln \frac{u_\nu(0)}{\hat u_\nu(0)}+\order(s^2).
\end{align}
Here, we used $\lim_{k\to\infty}u_\nu(ik)/\hat u_\nu(ik)=1$.
This reproduces the so-called Gelfand-Yaglom theorem \cite{Gelfand:1959nq}. 
Notice that $u_\nu(0)/\hat u_\nu(0)$ is well-defined in the limit of $r_\infty\to\infty$.
Plugging this into Eq.~\eqref{eq:lndet_from_zeta}, we obtain
\begin{equation}
    \left.\ln \frac{\det\mathcal M}{\det\widehat{\mathcal M}}\right|_{\zeta}=\sum_{\nu}\ab[d_\nu\ln R_\nu+\eta'_\nu(0)]-S'(0).\label{eq:zeta_function_renormalization}
\end{equation}
where
\begin{equation}
    R_\nu=\frac{\det\mathcal M_\nu}{\det\widehat{\mathcal M}_\nu}=\frac{u_\nu(0)}{\hat u_\nu(0)}=\lim_{r\to\infty}\frac{\psi_\nu(r)}{\hat \psi_\nu(r)}.\label{GY-theorem}
\end{equation}
Here, $\psi_\nu(r)$ and $\hat \psi_\nu(r)$ are the solutions of
\begin{align}
    \mathcal M_\nu\frac{\psi_\nu(r)}{r^{\frac{D}{2}-1}}=0,~\widehat{\mathcal M}_\nu\frac{\hat \psi_\nu(r)}{r^{\frac{D}{2}-1}}=0,\label{eq:GY}
\end{align}
with initial conditions
\begin{equation}
    \lim_{r\to0}\frac{\psi_\nu(r)}{r^{\nu}}=1,~\lim_{r\to0}\frac{\hat \psi_\nu(r)}{r^{\nu}}=1.
\end{equation}
Here, we factored out $r^{\frac{D}{2}-1}$ to make the differential equation slightly simpler,
\begin{equation}
    r^{\frac{D}{2}-1}\mathcal M_\nu\frac{1}{r^{\frac{D}{2}-1}}=-\partial_r^2-\frac{1}{r}\partial_r+\frac{\nu^2}{r^2}+m^2(r),
\end{equation}
which implies that $R_\nu$ is a function of $\nu^2$ only, instead of both the original angular momentum $l$ and the spacetime dimensions $D$.

The choice of $\eta_\nu(s)$ is not unique since we can always add any series that is convergent around $s=0$ to $\eta_\nu(s)$, and subtract its sum from $S(s)$. The simplest choice is the polynomials of $\nu^{i-2s}$ with $i\geq-1$, which has been obtained for $D\leq4$ in \cite{Dunne:2006ct} and will be extended to general dimensions in Section \ref{sec:large_nu}. Another choice is a series that is obtained by a WKB-like expansion of the heat kernel of the radial operator \cite{Hur:2008yg}. In Section \ref{sec:full_hke}, we present a new reference series that is compact and better approximates the UV behavior of functional determinants.

\section{Heat kernel expansion}\label{sec:hke}
Since the discussions in the following sections are based on the heat kernel expansion, we explain it briefly in this section.
For more details, see for example \cite{Wilk:1981vg}, \cite{Vassilevich:2003xt} and the references therein. 
 
The heat kernel of the operator $\mathcal M$ is the fundamental solution to the heat equation,
\begin{equation}
    (\partial_t+\mathcal M)K_{\mathcal M}(t,x',x)=0,
\end{equation}
where the differential operators in $\mathcal M$ operate on $x'$.
The initial condition is taken to be
\begin{equation}
    K_{\mathcal M}(0,x',x)=\delta^D(x'-x),
\end{equation}
where $\delta^D(x)$ is the $D$-dimensional delta distribution.

For the free operator $\widehat{\mathcal M}$, the heat kernel is explicitly given by
\begin{equation}
    K_{\widehat{\mathcal M}}(t,x',x)=\frac{e^{-\frac{(x'-x)^2}{4t}-\hat m^2t}}{(4\pi t)^{\frac{D}{2}}}.\label{eq:free_hk}
\end{equation}
For a more general $\mathcal M$, the heat kernel is formally written as
\begin{equation}
    K_{\mathcal M}(t,x',x)=\bra<x'|e^{-t\mathcal M}\ket|x>,
\end{equation}
where $\ket|x>$ is the basis in the position space.

The zeta function is related to the heat kernel via\footnote{
Another way of defining functional determinants is to use the proper-time representation, which is also based on the heat kernel:
\begin{equation*}
    \ln\det\mathcal M=-\int_{1/\Lambda^2}^\infty\frac{\odif{t}}{t}\mathcal K_{\mathcal M}(t).
\end{equation*}
This approach regularizes the UV divergence by introducing a UV cutoff $\Lambda$ on the integration domain, which is not required in the zeta function regularization thanks to the analyticity of the zeta function.}
\begin{equation}
    \zeta_{\mathcal M}(s)=\frac{1}{\Gamma(s)}\int_0^\infty\odif{t}t^{s-1}\mathcal K_{\mathcal M}(t),\label{eq:zeta_from_hk}
\end{equation}
where $\mathcal K_{\mathcal M}(t)$ is the trace of the heat kernel,
\begin{equation}
    \mathcal K_{\mathcal M}(t)=\tr e^{-t\mathcal M}=\int\odif[order=D]{x}K_{\mathcal M}(t,x,x).
\end{equation}
It should be noted that this expression diverges in the limit of infinite volume, analogous to the behavior of the zeta function, as can be readily seen from Eq.~\eqref{eq:free_hk}. For sufficiently fast decaying $\delta m^2$, the following quantity becomes finite:
\begin{equation}
    \mathcal K_{\mathcal M}(t)-\mathcal K_{\widehat{\mathcal M}}(t)=\int\odif[order=D]{x}\ab[K_{\mathcal M}(t,x,x)-K_{\widehat{\mathcal M}}(t,x,x)].
\end{equation}
Notice that this is finite for any $t>0$ since the UV contributions are suppressed exponentially. Instead, the UV behavior of the zeta function is encoded in the $t\sim0$ behavior of the heat kernel: the $t$ integral in Eq.~\eqref{eq:zeta_from_hk} converges for $\Re(s)>D/2$ because $\mathcal K_{\mathcal M}(t)$ behaves as $t^{-D/2}$ at a small $t$, as implied by Eq.~\eqref{eq:free_hk}.

The heat kernel expansion of $K_{\mathcal M}(t,x',x)$ around $K_{\widehat{\mathcal M}}(t,x',x)$ is the asymptotic expansion of
\begin{equation}
    K_{\mathcal M}(t,x',x)\approx K_{\widehat{\mathcal M}}(t,x',x)\sum_{a=0}^\infty t^ab_a(x',x),
\end{equation}
as $t\to+0$. Here the heat kernel has to be understood as a distribution in this limit.
We used $\approx$ to indicate that the expansion is asymptotic, meaning that the series does not necessarily converge. 

The heat kernel expansion can be understood more intuitively from a re-expansion of the Born series \cite{Wilk:1981vg}. Let us focus on the case where $x'=x$ for simplicity. The Born series refers to the series expansion of
\begin{align}
    \int_0^\infty\odif{t}K_{\mathcal M}(t,x,x)&=\braket<x|\frac{1}{-\partial^2+m^2}|x>\nonumber\\
    &\approx\sum_{n=0}^\infty(-1)^n\braket<x|\frac{1}{-\partial^2+\hat m^2}\ab[\delta m^2\frac{1}{-\partial^2+\hat m^2}]^n|x>.\label{eq:born_exp}
\end{align}
For each term of the expansion, the correlation is maximized when the positions $y_i$ of the inserted vertices $\delta m^2(y_i)$ coincide with $x$. Therefore, we further execute the derivative expansion with respect to the relative positions of $\delta m^2$'s. Then, we re-introduce the $t$ integral by rewriting the free Green's function with the free heat kernel.
The heat kernel coefficients are obtained by rearranging the series so that each order has the same power of $t$. Since $t$ has mass dimension $-2$, the term having $i$-number of $\delta m^2$ and $2j$-number of $\partial_\mu$ contributes to $b_{i+j}$, where $\partial_\mu$ only operates on $\delta m^2$.

The heat kernel coefficients $b_a(x',x)$ have been obtained in a closed form in \cite{Wilk:1981vg}. 
For $x'=x$, the first four coefficients are given by
\begin{align}
    b_0[\delta m^2]&=1,~b_1[\delta m^2]=-\delta m^2,~\label{eq:heat_kernel_coefficients1}
    b_2[\delta m^2]=\frac12\ab[(\delta m^2)^2-\frac13(\partial^2\delta m^2)],\\
    b_3[\delta m^2]&=-\frac16\ab[(\delta m^2)^3-\frac12(\partial_\mu\delta m^2)(\partial^\mu\delta m^2)-\delta m^2(\partial^2\delta m^2)+\frac{1}{10}(\partial^4\delta m^2)].\label{eq:heat_kernel_coefficients3}
\end{align}
Here, the derivatives operate only on $\delta m^2$ inside the parentheses.

Using the heat kernel coefficients, the zeta function is expanded as
\begin{align}
    \zeta_{\mathcal M}(s)-\zeta_{\widehat{\mathcal M}}(s)&\approx\sum_{a=1}^\infty\frac{1}{\Gamma(s)}\int_0^\infty\odif{t}t^{s+a-1}\tr e^{-t\widehat{\mathcal M}}b_a[\delta m^2].
\end{align}
A more convenient form is obtained by rewriting
\begin{equation}
    \mathcal K_{\mathcal M}(t)-\mathcal K_{\widehat{\mathcal M}}(t)=\int\odif[order=D]{x}\ab[K_{\mathcal M}(t,x,x)-K_{-\partial^2+z}(t,x,x)-K_{\widehat{\mathcal M}}(t,x,x)+K_{-\partial^2+z}(t,x,x)],
\end{equation}
and expanding $K_{\mathcal M}(t,x,x)$ and $K_{\widehat{\mathcal M}}(t,x,x)$ around $(-\partial^2+z)$. Here, $z$ is a real parameter satisfying $z\geq0$ to avoid the negative and zero modes of $(-\partial^2+z)$. We obtain
\begin{align}
    \zeta_{\mathcal M}(s)-\zeta_{\widehat{\mathcal M}}(s)&\approx\sum_{a=1}^\infty\frac{1}{\Gamma(s)}\int_0^\infty\odif{t}t^{s+a-1}\tr e^{-t(-\partial^2+z)}\ab(b_a[m^2-z]-b_a[\hat m^2-z]).\label{eq:convenient_form}
\end{align}
As we have mentioned earlier, $z=0$ is allowed as the Laplace operator does not have a zero mode. However, we will see in Section \ref{sec:full_hke} that there appear artificial infrared (IR) divergences for $z=0$, which must be treated carefully if the IR contributions are relevant to the calculation.

\section{Large angular momentum subtraction}\label{sec:large_nu}
In \cite{Dunne:2006ct}, the reference series $\eta_\nu(s)$ is obtained by the large angular momentum expansion of $u_\nu(ik)$ in Eq.~\eqref{eq:zeta_robust}.
However, this analysis becomes significantly difficult for $D>5$, where we need higher orders of the expansion.
In this section, we present a much simpler procedure to identify the reference series, and explicitly obtain higher-order terms that suffice to regularize up to 13-dimensional functional determinants.

Our proposal is to evaluate the trace of Eq.~\eqref{eq:convenient_form} in the angular momentum space, and expand it at a large angular momentum. We choose $z=0$ to simplify the analysis, ignoring the IR divergences that appear in small angular momenta.
Let us first introduce two basis sets in the angular momentum space:
\begin{equation}
    1=\sum_{\nu,\chi}\int_0^\infty\odif{r}r^{D-1}\ket|\nu,\chi,r>\bra<\nu,\chi,r|=\sum_{\nu,\chi}\int_0^\infty\odif{\lambda}\ket|\nu,\chi,\lambda>\bra<\nu,\chi,\lambda|,\label{eq:complete_set_angular}
\end{equation}
where
\begin{align}
    \braket<x|\nu,\chi,r>&=\delta(|x|-r)Y_{(\nu+1-D/2)\chi}(\hat x),\\
    \braket<x|\nu,\chi,\lambda>&=\sqrt{\lambda}|x|^{\frac{2-D}{2}}J_\nu(\lambda |x|)Y_{(\nu+1-D/2)\chi}(\hat x).
\end{align}
Here, $J_\nu(z)$ is the Bessel function of the first kind, and the basis $\ket|\nu,\chi,\lambda>$ diagonalizes the Laplace operator as
\begin{equation}
    -\partial^2\ket|\nu,\chi,\lambda>=\lambda^2\ket|\nu,\chi,\lambda>.
\end{equation}

A naive expression of the trace in the angular momentum space would be
\begin{equation}
    \tr e^{-t(-\partial^2)}\ab(b_a[m^2]-b_a[\hat m^2])=\sum_{\nu,\chi}\tr P_{\nu\chi}e^{-t(-\partial^2)}\ab(b_a[m^2]-b_a[\hat m^2]),
\end{equation}
where the projection operator is defined by
\begin{equation}
    P_{\nu\chi}=\int_0^\infty\odif{\lambda}\ket|\nu,\chi,\lambda>\bra<\nu,\chi,\lambda|.
\end{equation}
Since the trace is independent of $\chi$ due to spherical symmetry, we will use the notation $P_{\nu}$ for a representative $\chi$, and multiply the degeneracy factor $d_\nu$ in the remainder of the article.

Keeping in mind that we are working in the convergent regime $\Re(s)>D/2-1$, we interchange the $t$ integral and the sum as
\begin{align}
    \zeta_{\mathcal M}(s)-\zeta_{\widehat{\mathcal M}}(s)&\approx\sum_{a=1}^\infty\frac{1}{\Gamma(s)}\int_0^\infty\odif{t}t^{s+a-1}\sum_\nu d_\nu\tr P_\nu e^{-t(-\partial^2)}\ab(b_a[m^2]-b_a[\hat m^2])\nonumber\\
    &=\sum_{a=1}^\infty\frac{1}{\Gamma(s)}\sum_\nu d_\nu\int_0^\infty\odif{t}t^{s+a-1}\tr P_\nu e^{-t(-\partial^2)}\ab(b_a[m^2]-b_a[\hat m^2]).
\end{align}
Now, the question is whether the coefficients in the large $\nu$ expansion of the above expression coincide with those in Eq.~\eqref{eq:zeta_robust}.
Unfortunately, they do not, even though $\nu$ itself is a well-defined conserved quantity.
This is caused by the use of the heat kernel expansion, which is an expansion in the momentum space and does not track the angular momentum explicitly. Therefore, we have to adjust the coefficients so that they become compatible with the angular momentum expansion.
Such an adjustment may be possible because total derivative terms can be freely added to $b_a$ without affecting the value of the trace for $\Re(s)>D/2-1$, while still altering the coefficients in the large $\nu$ expansion.

Let us elaborate on how it is possible to change the large $\nu$ behavior without affecting the total sum.
One such example is
\begin{equation}
    \sum_{\nu=1}^{N}\ab[\frac{1}{\zeta(s)\nu^s}-\frac{1}{\zeta(s+1)\nu^{s+1}}].\label{eq:example_div}
\end{equation}
As $N\to\infty$, it converges to zero for $\Re(s)>1$, and thus the corresponding analytic function is zero. However, the two terms have different large $\nu$ behavior. This is exactly what happens when we add total derivatives: for an arbitrary total derivative $\partial_\mu A^\mu$, we have $\tr e^{-t(-\partial^2)}(\partial_\mu A^\mu)=0$, but $\tr P_\nu e^{-t(-\partial^2)}(\partial_\mu A^\mu)$ is generally non-zero.

We thus ask the next question whether there exist total derivatives that make the large $\nu$ behavior agree with that of \eqref{eq:zeta_robust}. Although it is difficult to rigorously prove that such total derivatives exist for all $b_a$, we propose a systematic way to determine them in Appendix \ref{apx:matching}. We found that we do have such total derivatives at least up to $b_6$ and up to $\order(s)$, and the results up to $a=5$ are
\begin{align}
    B_1^{z=0}[m^2]&=-m^2,\label{eq:large_b_from}\\
    B_2^{z=0}[m^2]&=\frac12(m^2)^2,\\
    B_3^{z=0}[m^2]&=-\frac16\ab[(m^2)^3+\frac12(\partial_\mu m^2)(\partial^\mu m^2)],\label{eq:large_b_3}\\
    B_4^{z=0}[m^2]&=\frac{1}{24}\left[(m^2)^4+2 m^2(\partial_\mu m^2)(\partial^\mu m^2)\right.\\
    &\hspace{8ex}\left.+\frac{4}{15}(\partial_\mu\partial_\nu m^2)(\partial^\mu\partial^\nu m^2)+\frac{1}{15}(\partial^\mu m^2)(\partial_\mu\partial^2 m^2)\right],\label{eq:large_b_to}
\end{align}
and
\begin{align}
    B_5^{z=0}[m^2]&=-\frac{1}{120}\left[(m^2)^5+5(m^2)^2(\partial_\mu m^2)(\partial^\mu m^2)+\frac54 m^2(\partial_\mu\partial_\nu m^2)(\partial^\mu\partial^\nu m^2)\right.\nonumber\\
    &\hspace{10ex}+\frac14 m^2(\partial_\mu m^2)(\partial^2\partial^\mu m^2)+\frac{13}{6}(\partial_\mu  m^2)(\partial_\nu  m^2)(\partial^\mu\partial^\nu m^2)\nonumber\\
    &\hspace{10ex}+\frac{1}{8}(\partial_\mu  m^2)(\partial^\mu  m^2)(\partial^2 m^2)+\frac{1}{7}(\partial_\mu\partial_\nu\partial_\rho m^2)(\partial^\mu\partial^\nu\partial^\rho m^2)\nonumber\\
    &\hspace{10ex}+\frac{5}{56}(\partial_\mu\partial_\nu m^2)(\partial^\mu\partial^\nu\partial^2 m^2)+\frac{1}{112}(\partial_\mu\partial^2 m^2)(\partial^\mu\partial^2 m^2)\nonumber\\
    &\hspace{10ex}\left.+\frac{1}{112}(\partial_\mu m^2)(\partial^\mu\partial^4 m^2)\right].
\end{align}
They do not depend on $D$ and are unique up to $a=5$ under the assumption that the total derivatives are written only with $m^2$ and its derivatives. We also give $B_6^{z=0}[m^2]$ in Appendix \ref{apx:matching}, which is found to be not unique and has two arbitrary parameters.

We observe that $B_a^{z=0}[m^2]$ is much simpler than the original heat kernel coefficient $b_a[m^2]$, {\it e.g.} $B_5^{z=0}[m^2]$ has 10 terms and $b_5[m^2]$ has 16 terms. In particular, the terms like $\partial^2\delta m^2,\partial^4\delta m^2,\cdots$ disappear in $B_a^{z=0}[m^2]$. Such total derivative terms are indeed artifacts of the heat kernel expansion because they originate from the relative position of $\delta m^2$ with respect to the end points $x$ of $K_{\mathcal M}(t,x,x)$, which has no physical meaning once we take the trace.

Using these coefficients, we obtain the expansion of zeta function having the same large $\nu$ behavior as\footnote{Since we have not shown the existence of $B_a^{z=0}[m^2]$ for $a>6$, the most conservative definition is $B_a^{z=0}[m^2]=b_a[m^2]$ for $a>6$, which is enough for the regularization in up to 13 dimensions.}
\begin{align}
    \zeta_{\mathcal M}(s)-\zeta_{\widehat{\mathcal M}}(s)&\approx\sum_{a=1}^\infty\frac{1}{\Gamma(s)}\sum_\nu d_\nu\int_0^\infty\odif{t}t^{s+a-1}\tr P_\nu e^{-t(-\partial^2)}\ab(B_a^{z=0}[m^2]-B_a^{z=0}[\hat m^2])\nonumber\\
    &\hspace{3ex}+\sum_\nu d_\nu \Delta_\nu(s).
\end{align}
Here, $\sum_\nu\Delta_\nu(s)$ absolutely converges to zero for $\Re(s)>D/2-1$ and $\Delta_\nu(s)=\order(s^2)$.
It is introduced only to correct the large $\nu$ behavior at $\order(s^2)$ if necessary, and is irrelevant for the calculation of determinants. Since one can show $\Delta_\nu(s)=0$ around $s=0$ up to $B_2^{z=0}[m^2]$ following \cite{Dunne:2006ct},
it might be possible to prove $\Delta_\nu(s)=0$ in general. Nevertheless, this is beyond the scope of this paper, and we simply omit $\Delta_\nu(s)$ in the following to save space.

The trace is evaluated by inserting complete sets as
\begin{align}
    &\tr P_{\nu} e^{-t(-\partial^2)}\ab(B_a^{z=0}[m^2]-B_a^{z=0}[\hat m^2])\nonumber\\
    &=\int_0^\infty\odif{\lambda}\int_0^\infty\odif{r}r^{D-1}\bra<\nu,\chi,\lambda|e^{-\lambda^2t}\ket|\nu,\chi,r>\bra<\nu,\chi,r|\ab(B_a^{z=0}[m^2]-B_a^{z=0}[\hat m^2])\ket|\nu,\chi,\lambda>\nonumber\\
    &=\int_0^\infty\odif{r}r\ab(B_a^{z=0}[m^2](r)-B_a^{z=0}[\hat m^2](r))\int_0^\infty\odif{\lambda}e^{-\lambda^2t}\lambda J_\nu^2(\lambda r)\nonumber\\
    &=\int_0^\infty\odif{r}r\ab(B_a^{z=0}[m^2](r)-B_a^{z=0}[\hat m^2](r))\frac{e^{-\frac{r^2}{2t}}}{2t}I_\nu\ab(\frac{r^2}{2t}),
\end{align}
where $I_\nu(z)$ is the modified Bessel function of the first kind.

Executing the $t$ integral, we obtain
\begin{align}
    \zeta_{\mathcal M}(s)-\zeta_{\widehat{\mathcal M}}(s)&\approx\sum_{a=1}^\infty\frac{1}{\Gamma(s)}\sum_\nu d_\nu\Xi_{a,\nu}(s)\int_0^\infty\odif{r}r^{2(a+s)-1}\ab(B_a^{z=0}[m^2](r)-B_a^{z=0}[\hat m^2](r)),\label{eq:before_large_nu}
\end{align}
where
\begin{align}
    \Xi_{a,\nu}(s)=\frac{1}{2}\int_0^\infty\odif{t'}t'^{s+a-2}e^{-\frac{1}{2t'}}I_\nu\ab(\frac{1}{2t'})=\frac{\Gamma\ab(a+s-\frac12)}{2\sqrt{\pi}}\frac{\Gamma\ab(\nu-a+1-s)}{\Gamma\ab(\nu+a+s)}.\label{eq:t-integral_xi}
\end{align}
Notice that we rescaled the integration variable, $t=t'r^2$, to separate the $r$ dependence.

Finally, we expand the $\nu$ dependent part at $\nu\to\infty$,
\begin{align}
    d_\nu\Xi_{a,\nu}(s)&\approx\sum_{i=0}^\infty c_{a}^{(2i)}(s)\nu^{D-1-2(i+a)-2s}.\label{eq:large_nu_exp}
\end{align}
Notice that $c_a^{(2i+1)}(s)=0$ because $d_\nu\Xi_{a,\nu}(s)$ is analytic with respect to $\nu$ and is an odd (even) function of $\nu$ for even (odd) dimensions.

Since we have identified the divergent contributions, we define a reference series as
\begin{align}
    \eta_\nu^{\rm LAM}(s)&=\sum_{a=1}^{a_{\rm max}}\sum_{i=0}^{a_{\max}-a}\nu^{D-1-2(i+a)-2s}\frac{c_a^{(2i)}(s)}{\Gamma(s)}\int_0^\infty\odif{r}r^{2(a+s)-1}\ab(B_a^{z=0}[m^2](r)-B_a^{z=0}[\hat m^2](r)).\label{eq:ref_lam}
\end{align}
Here, $a_{\max}$ is an integer satisfying $a_{\max}>D/2-1$.
Notice that Eq.~\eqref{eq:ref_lam} has an artificial IR divergence at $\nu=0$ in two dimensions. Since we are not interested in small $\nu$ contributions, we simply define $\eta_0^{\rm LAM}(s)=0$ for $D=2$.

Now, the reference series, Eq.~\eqref{eq:ref_lam}, is a simple function of $s$, and its sum $S^{\rm LAM}(s)=\sum_\nu\eta_\nu^{\rm LAM}(s)$ is finite for $\Re(s)>D/2-1$.
The analytic continuation of $S^{\rm LAM}(s)$ is easily obtained because we know $\sum_\nu\nu^{-s}=\zeta\ab(s,D/2-1)$,
where $\zeta(s,a)$ is the Hurwitz zeta function. Notice that the Hurwitz zeta functions in $S^{\rm LAM}(s)$ have a pole at $s=0$ in even dimensions like the dimensional regularization, which generates logarithmic terms. However, the zeta function has an overall $1/\Gamma(s)\simeq s$ factor, which eliminates the pole eventually. This is how the pole at $s=0$ disappears in the zeta function regularization.

One can easily check that Eq.~\eqref{eq:ref_lam} with $a_{\max}=2$ reproduces the results of \cite{Dunne:2006ct}.
For example, the result in four dimensions is
\begin{align}
    -\eta'^{\rm LAM}_\nu(0)=\frac{\nu}{2}\int_0^\infty\odif{r}r(m^2-\hat m^2)-\frac{1}{8\nu}\int_0^\infty\odif{r}r^{3}(m^4-\hat m^4),
\end{align}
and
\begin{align}
    -S'^{\rm LAM}(0)=-\frac{1}{8}\int_0^\infty\odif{r}r^{3}(m^4-\hat m^4)\ab(1+\gamma+\ln\frac{r}{2}),
\end{align}
which is in agreement with \cite{Dunne:2006ct}.

We emphasize that our method facilitates the computation of the reference series for larger $a$ (with explicit formulas provided up to $a=6$), which would be prohibitively complex using the original approach.

\section{Full heat kernel coefficient subtraction}\label{sec:full_hke}
In the previous section, we expanded $d_\nu\Xi_{a,\nu}(s)$ at a large $\nu$ to obtain the reference series of \cite{Dunne:2006ct}.
However, we could subtract the entire heat kernel coefficients, which is preferable because they are well-defined quantities beyond the angular momentum expansion.
Thus, let us first consider a reference series given by
\begin{align}
    \eta^{\rm HKC}_\nu(z=0;s)&=\sum_{a=1}^{a_{\rm max}}\frac{d_\nu\Xi_{a,\nu}(s)}{\Gamma(s)}\int_0^\infty\odif{r}r^{2(a+s)-1}\ab(B_a^{z=0}[m^2](r)-B_a^{z=0}[\hat m^2](r)).\label{eq:hke_z0}
\end{align}
However, it has artificial IR divergences at $s=0$ for $\nu\leq a-1$, which can be seen from the large $t$ behavior of Eq.~\eqref{eq:t-integral_xi}. This is caused by the massless Green's functions used in the heat kernel expansion, and we will come back to this point later in this section.

A simple way to avoid these IR divergences is to adopt a non-zero mass, $z>0$. Following the same procedure as in the previous section, we obtain
\begin{align}
    \zeta_{\mathcal M}(s)-\zeta_{\widehat{\mathcal M}}(s)&\approx\sum_{a=1}^\infty\frac{1}{\Gamma(s)}\sum_\nu d_\nu\int_0^\infty\odif{r}r\ab(B_a^{z}[m^2-z](r)-B_a^{z}[\hat m^2-z](r))\mathcal I_{a+s,\nu}(z;r),\label{eq:convenient_form_angular}
\end{align}
where
\begin{align}
    \mathcal I_{a,\nu}(z;r)&=\frac12\int_0^\infty\odif{t}t^{a-2}e^{-\frac{r^2}{2t}-zt}I_\nu\ab(\frac{r^2}{2t}).\label{eq:integral_hke}
\end{align}
Here, $B_a^{z}[m^2]$ can differ from $B_a^{z=0}[m^2]$ by additional total derivatives $\Delta B_a^z[m^2]$ that depend on $z$:
\begin{equation}
    B_a^{z}[m^2]=B_a^{z=0}[m^2]+\Delta B_a^z[m^2].
\end{equation}

We determine $\Delta B_a^z[m^2]$ by expanding $\mathcal I_{s,\nu}(z;r)$ at a large $\nu$ and comparing it with Eq.~\eqref{eq:before_large_nu}. Since we know that the insertion of the mass $z$ to a Feynman diagram decreases the order of divergence, we expand the integral for a small $z$ as\footnote{Precisely speaking, this expansion is valid up to $\order(z^{\nu})$.
Since we are interested in the first few terms at a large $\nu$, this is not an issue at all.}
\begin{align} 
    \mathcal I_{a,\nu}(z;r)&\approx\frac12\sum_{n=0}^\infty\frac{(-z)^n}{n!}\int_0^\infty\odif{t}t^{a+n-2}e^{-\frac{r^2}{2t}}I_\nu\ab(\frac{r^2}{2t})\nonumber\\
    &=\sum_{n=0}^\infty\frac{(-z)^n}{n!}\Xi_{a+n,\nu}(0)r^{2(a+n-1)}.\label{eq:small_mhat_exp}
\end{align}
This helps us to compare the large $\nu$ behavior.
Matching the coefficients of $\Xi_a(s)$ with those of \eqref{eq:before_large_nu}, we obtain
\begin{align}
    \Delta B_1^z[m^2]&=\Delta B_2^z[m^2]=\Delta B_3^z[m^2]=\Delta B_4^z[m^2]=0,\\
    \Delta B_5^z[m^2]&=-\frac{1}{120}\frac{-z}{12}\ab[(\partial_\mu\partial_\nu m^2)(\partial^\mu\partial^\nu m^2)+(\partial^\mu m^2)(\partial_\mu\partial^2 m^2)].
\end{align}
We also provide $\Delta B_6^z[m^2]$ in Appendix \ref{apx:matching}.

It is noteworthy that the corrections $\Delta B_i^z[m^2]$ vanish for $i=1, \dots, 4$. For $i=1,2,3$, this result follows naturally from dimensional analysis and rotational symmetry constraints: While $z\partial^2 m^2$ could contribute for $i=3$, it appears to be absent by the same mechanism that excludes terms of the form $\partial^2 m^2,\partial^4m^2,\cdots$ in $B_i^{z=0}[m^2]$.
The vanishing of the correction for $i=4$ is more subtle, as there is a possible total derivative term $\partial_\mu[m^2\partial^\mu m^2]$ but is found to be absent.

We define our reference series as
\begin{align}
    \eta^{\rm HKC}_\nu(z;s)&=\sum_{a=1}^{a_{\rm max}}\frac{d_\nu}{\Gamma(s)}\int_0^\infty\odif{r}r\ab(B_a^{z}[m^2-z](r)-B_a^{z}[\hat m^2-z](r))\mathcal I_{a+s,\nu}(z,r).\label{eq:eta_hke}
\end{align}
For $s=0$, the integral of Eq.~\eqref{eq:integral_hke} gives a simple formula,
\begin{align}
    \mathcal I_{a,\nu}(z;r)&=\ab(-\pdv{}{z})^{a-1}I_\nu(\sqrt{z} r)K_\nu(\sqrt{z} r),
\end{align}
where $K_\nu(z)$ is the modified Bessel function of the second kind. For $a=1$, this is the well-known free radial Green's function. For $a>1$, one can get explicit formulas using $\odif{I_\nu(z)}/\odif{z}=(I_{\nu+1}(z)+I_{\nu-1}(z))/2$, $\odif{K_\nu(z)}/\odif{z}=-(K_{\nu+1}(z)+K_{\nu-1}(z))/2$, $I_{-\nu}(z)=I_\nu(z)$, and $K_{-\nu}(z)=K_\nu(z)$. As one can easily check, the artificial IR divergences for $\nu\leq a-1$ are regularized by $z$.

Let us move on to the computation of the sum $S^{\rm HKC}(s)$.
One of the benefits of using the entire heat kernel coefficients is that $S^{\rm HKC}(s)$ can be directly evaluated in the momentum space. To make the discussion parallel to that in the angular momentum space, we introduce complete sets,
\begin{equation}
    1=\int\odif[order=D]{x}\ket|x>\bra<x|=\int\frac{\odif[order=D]{k}}{(2\pi)^D}\ket|k>\bra<k|,
\end{equation}
where
\begin{equation}
    -\partial^2\ket|k>=k^2\ket|k>,~\braket<x|k>=e^{ikx}.
\end{equation}
Inserting these into the trace, we obtain
\begin{align}
    S^{\rm HKC}(z;s)&=\frac{1}{\Gamma(s)}\sum_{a=1}^{a_{\max}}\int_0^\infty\odif{t}t^{s+a-1}\int\odif[order=D]{x}\int\frac{\odif[order=D]{k}}{(2\pi)^D} \bra<k|e^{-t(-\partial^2+z)}\ket|x>\nonumber\\
    &\hspace{3ex}\times\braket<x|\ab(B_a^{z}[m^2-z](x)-B_a^{z}[\hat m^2-z](x))|k>\nonumber\\
    &=\frac{\Gamma(a+s)}{\Gamma(s)}\sum_{a=1}^{a_{\max}}\int\odif[order=D]{x}\ab(B_a^{z}[m^2-z](x)-B_a^{z}[\hat m^2-z](x))\int\frac{\odif[order=D]{k}}{(2\pi)^D}\frac{1}{(k^2+z)^{a+s}}\nonumber\\
    &=\sum_{a=1}^{a_{\max}}\frac{\Gamma\ab(a+s-\frac{D}{2})}{2^{D-1}\Gamma(s)\Gamma\ab(\frac{D}{2})}z^{\frac{D}{2}-a-s}\int\odif{r}r^{D-1}\ab(B_a^{z}[m^2-z](r)-B_a^{z}[\hat m^2-z](r)).\label{eq:s_hke}
\end{align}
Here we interchanged the $t$ integral and the momentum integrals, which is justified for $\Re(s)>D/2-1$. Since $S^{\rm HKC}(s)$ is explicitly analytic in $s$, its analytic continuation to $\Re(s)\leq D/2-1$ is trivial. We also clearly see that all the IR divergences are regularized by $z$.

A different choice of $z$ results in a different reference series, and a detailed comparison will be given in Section \ref{sec:vacuum_decay}. We will see that the series is similar to the one with the large angular momentum expansion when $z=0$, and is similar to the one in the Feynman diagrammatic approach when $z=\hat m^2$. No matter which value is chosen, the sum of the difference $\eta^{\rm HKC}_\nu(z;s)-\eta^{\rm HKC}_\nu(z';s)$ is finite and is exactly canceled by $S^{\rm HKC}(z;s)-S^{\rm HKC}(z';s)$ in the final result.

Now, let us go back to Eq.~\eqref{eq:hke_z0}.
As seen from Eqs.~\eqref{eq:eta_hke} and \eqref{eq:s_hke}, we cannot take $z\to\hat m^2$ around $s=0$ because of the artificial IR divergences. However, these IR divergences should cancel between $\sum_\nu\eta'^{\rm HKC}_{\nu}(z\to0;0)$ and $S'^{\rm HKC}(z\to0;0)$.
This can be seen from the independence of Eq.~\eqref{eq:zeta_function_renormalization} on the choice of $z$, as well as from the absence of IR divergences for $z>0$.

We can also see the cancellation explicitly. For example, in four dimensions, a log divergence appears at the lowest angular momentum $\nu=1$,
\begin{align}
    -\eta'^{\rm HKC}_{1}(z;0)&=\frac12\int\odif{r}r(m^2-\hat m^2)\nonumber\\
    &\hspace{3ex}+\frac{1}{8}\int \odif{r}r^3(m^4-\hat m^4)\ab(\frac14+\gamma+\ln\frac{r}{2}+\frac12\ln z)+\order(z),
\end{align}
which is indeed canceled by
\begin{equation}
    -S'^{\rm HKC}(z;0)=\frac{1}{16}\int\odif{r}r^3(m^4-\hat m^4)\ln z.
\end{equation}

There are two equivalent approaches to handle the IR divergences, and the choice between them can be made based on implementation preferences.

The first approach is to pre-include the IR divergences in the add-back term.
We define
\begin{align}
    \tilde\eta^{\rm HKC}_\nu(s)&=\sum_{a=1}^{a_{\rm max}}\frac{d_\nu\widetilde \Xi_{a,\nu}(s)}{\Gamma(s)}\int_0^\infty\odif{r}r^{2(a+s)-1}\ab(B_a^{z=0}[m^2](r)-B_a^{z=0}[\hat m^2](r)),\label{eq:tilde_eta_hke}
\end{align}
where
\begin{equation}
    \widetilde \Xi_{a,\nu}(s)=
    \begin{cases}
        0&\nu\leq\bar\nu_a\\
        \Xi_{a,\nu}(s)&\nu> \bar\nu_a
    \end{cases},
\end{equation}
and
\begin{equation}
    \bar\nu_a=
    \begin{cases}
        a-1&({\rm even~dimensions})\\
        a-\frac32&({\rm odd~dimensions})
    \end{cases}.
\end{equation}
Then, the terms having $\nu\leq\bar \nu_a$ are moved to the add-back term as
\begin{align}
    \tilde S^{\rm HKC}(s)&=\lim_{z\to+0}\ab[S^{\rm HKC}(z;s)-\eta^{\rm HKC}_{\rm low}(z;s)],\\
    \eta^{\rm HKC}_{\rm low}(z;s)&=\sum_{a=1}^{a_{\rm max}}\sum_{\nu}^{\bar\nu_a}\frac{d_\nu}{\Gamma(s)}\int_0^\infty\odif{r}r\ab(B_a^{z}[m^2-z](r)-B_a^{z}[\hat m^2-z](r))\mathcal I_{a+s,\nu}(z,r).
\end{align}
One may also increase $\bar\nu_a$ to include more terms in the add-back term because the lower $\nu$ contributions with $z=0$ are typically large.

The second approach is to regularize the IR divergences using the zeta function regularization. 
The artificial IR divergences are handled through analytic continuation in $s$, in a manner analogous to the treatment of UV divergences, though the regularization occurs in different domains.
This is much simpler because we can use analytically continued values of Eq.~\eqref{eq:hke_z0} for all $\nu$, and also because the add-back term is zero, $S^{\rm HKC}(0;s)=0$. Here, we note that $\int \odif[order=D]{k}k^{-2(a+s)}=0$ due to the absence of dimensionful parameters\footnote{For a more rigorous discussion, one needs to separately execute analytic continuation for the IR divergence and the UV divergence.}.
It is noteworthy that this reference series effectively subtracts a net zero from the determinant. This occurs because the UV divergent contributions and the artificial IR divergences precisely cancel in $S^{\rm HKC}(0;s)$.
This cancellation implies that the terms present in $S'^{\rm HKC}(z>0;0)$, including those proportional to $\ln r$, are effectively incorporated within the small $\nu$ contributions of $\eta^{\rm HKC}_\nu(0;s)$.

\section{Dimensional regularization}\label{sec:dimreg}
In the previous sections, we have only discussed the zeta function regularization. It is slightly different from the dimensional regularization, on which the coupling constants in the $\overline{\rm MS}$ scheme are defined. In particular, the zeta function regularization does not generate the pole and the renormalization scale\footnote{It is possible to restore the renormalization scale in the zeta function regularization by rescaling the fluctuation operator as $\zeta_{\mathcal M/\mu^2}(s)$ \cite{Dunne:2006ct}.}, which appear in the dimensional regularization in even dimensions.
Although the relation between the two regularization schemes has been discussed from a general perspective \cite{Bytsenko:2003tu} and explicitly in four dimensions \cite{Baratella:2025dum}, it is instructive to derive the conversion formula for completeness.

\begin{figure}[t]
    \centering
    \includegraphics[width=0.4\textwidth]{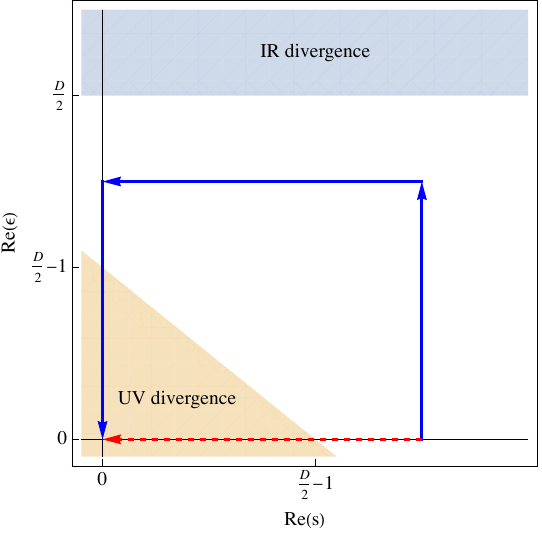}
    \caption{Convergent region of $S^{\rm HKC}(0;s;\varepsilon)$. In the white region, the momentum integral is absolutely convergent. The shaded regions are IR/UV divergent and are defined only through analytic continuation of either $s$ or $\varepsilon$. The red dashed line corresponds to the zeta function regularization and the blue solid line corresponds to the dimensional regularization. }
    \label{fig:dimreg}
\end{figure}

The dimensional regularization is defined through the analytic continuation of the spacetime dimensions of the momentum integral from $D$ to $D-2\varepsilon$. In the convergent region $\Re(s)>D/2-1$, we generalize Eq.~\eqref{eq:s_hke} with $z=\hat m^2$ as
\begin{align}
    S^{\rm HKC}(\hat m^2;s;\varepsilon)&=\frac{\Gamma(a+s)}{\Gamma(s)}\sum_{a=1}^{a_{\max}}\int\odif[order=D]{x}B_a[\delta m^2](x)\int\frac{\odif[order=D-2\varepsilon]{k}}{(2\pi)^{D-2\varepsilon}}\frac{\mu^{2\varepsilon}}{(k^2+\hat m^2)^{a+s}}\nonumber\\
    &=\sum_{a=1}^{a_{\max}}\frac{\Gamma\ab(a+s-\frac{D}{2}+\varepsilon)}{(4\pi)^{\frac{D}{2}-\varepsilon}\Gamma(s)}\frac{\mu^{2\varepsilon}}{(\hat m^2)^{a+s-\frac{D}{2}+\varepsilon}}\int\odif[order=D]{x}B_a[\delta m^2](x),
\end{align}
where $\mu$ is the renormalization scale. Since we are only interested in divergent terms, we take $a_{\max}=D/2$ for even dimensions and $a_{\max}=(D-1)/2$ for odd dimensions.
Here, $\delta m^2(x)$ is not necessarily spherically symmetric.

The momentum integral is absolutely convergent for the white region in Fig.~\ref{fig:dimreg}, and thus its value is unique.
For $D/2>\Re(\varepsilon)>D/2-1$, we can take $s\to0$ safely, and obtain an analytic function of $\varepsilon$ as
\begin{align}
    -\pdv{S^{\rm HKC}}{s}(\hat m^2;0;\varepsilon)&=-\sum_{a=1}^{a_{\max}}\frac{\Gamma\ab(a-\frac{D}{2}+\varepsilon)}{(4\pi)^{\frac{D}{2}-\varepsilon}}\frac{\mu^{2\varepsilon}}{(\hat m^2)^{a-\frac{D}{2}+\varepsilon}}\int\odif[order=D]{x}B_a[\delta m^2](x).
\end{align}

The result in the dimensional regularization is obtained by analytically continuing it to $\varepsilon\simeq0$.
For odd dimensions, both regularization schemes give the same result,
\begin{align}
    -\lim_{s\to0}\pdv{S^{\rm HKC}}{s}(\hat m^2;s;0)&=-\lim_{\varepsilon\to0}\pdv{S^{\rm HKC}}{s}(\hat m^2;0;\varepsilon)\nonumber\\
    &=-\sum_{a=1}^{(D-1)/2}\frac{\Gamma\ab(a-\frac{D}{2})}{(4\pi)^{\frac{D}{2}}}\frac{1}{(\hat m^2)^{a-\frac{D}{2}}}\int\odif[order=D]{x}B_a[\delta m^2](x).
\end{align}

However, for even dimensions, the two regularization schemes give slightly different results. For the zeta function regularization, we have
\begin{align}
    -\pdv{S^{\rm HKC}}{s}(\hat m^2;s;0)&=-\sum_{a=1}^{D/2}\frac{(\hat m^2)^{\frac{D}{2}-a}}{(4\pi)^{\frac{D}{2}}}\frac{(-1)^{\frac{D}{2}-a}}{\ab(\frac{D}{2}-a)!}\ab(H_{\frac{D}{2}-a}-\ln\hat m^2)\int\odif[order=D]{x}B_a[\delta m^2](x)+\order(s),
\end{align}
where $H_n$ is the harmonic number. On the other hand, the dimensional regularization gives
\begin{align}
    -\pdv{S^{\rm HKC}}{s}(\hat m^2;0;\varepsilon)&=-\sum_{a=1}^{D/2}\frac{(\hat m^2)^{\frac{D}{2}-a}}{(4\pi)^{\frac{D}{2}}}\frac{(-1)^{\frac{D}{2}-a}}{\ab(\frac{D}{2}-a)!}\ab(\frac{1}{\bar\varepsilon}+H_{\frac{D}{2}-a}-\ln\frac{\hat m^2}{\mu^2})\int\odif[order=D]{x}B_a[\delta m^2](x)+\order(\varepsilon),
\end{align}
where $1/\bar\varepsilon=1/\varepsilon-\gamma+\ln4\pi$.

To summarize, the ratio of the functional determinants in the dimensional regularization is expressed as
\begin{equation}
    \left.\ln \frac{\det\mathcal M}{\det\widehat{\mathcal M}}\right|_{\rm DR}=\sum_{\nu}\ab[d_\nu\ln R_\nu+\eta'_\nu(0)]-S'(0)+\delta_{\rm DR},\label{eq:dimreg-conversion}
\end{equation}
where $\delta_{\rm DR}=0$ for odd dimensions, and
\begin{equation}
    \delta_{\rm DR}=-\sum_{a=1}^{D/2}\frac{(\hat m^2)^{\frac{D}{2}-a}}{(4\pi)^{\frac{D}{2}}}\frac{(-1)^{\frac{D}{2}-a}}{\ab(\frac{D}{2}-a)!}\ab(\frac{1}{\bar\varepsilon}+\ln\mu^2)\int\odif[order=D]{x}B_a[\delta m^2](x),
\end{equation}
for even dimensions. This holds also for $\hat m^2=0$ since both regularization schemes have the same $\ln \hat m^2$ dependence.
Notice that $\delta_{\rm DR}$ is independent of the choice of the reference series, as is apparent from Eq.~\eqref{eq:dimreg-conversion}.

Finally, let us explain why the limit of $(s,\varepsilon)\to(0,0)$ is not uniquely determined. Around $s=\varepsilon=0$, the function is expanded as
\begin{align}
    -\pdv{S^{\rm HKC}}{s}(\hat m^2;s;\varepsilon)&=-\sum_{a=1}^{D/2}\frac{(\hat m^2)^{\frac{D}{2}-a}}{(4\pi)^{\frac{D}{2}}}\frac{(-1)^{\frac{D}{2}-a}}{\ab(\frac{D}{2}-a)!}\int\odif[order=D]{x}B_a[\delta m^2](x)(1+\varepsilon\ln\mu^2+\varepsilon\ln4\pi)\nonumber\\
    &\hspace{3ex}\times\ab[\ab(\frac{1}{s+\varepsilon}+H_{\frac{D}{2}-a}-\gamma-\ln\hat m^2)(1+2s\gamma)-\frac{s(1+s\gamma)}{(s+\varepsilon)^2}]+\order(s,\varepsilon).
\end{align}
It has terms like $s/(s+\varepsilon)$ and $\varepsilon/(s+\varepsilon)$, which either goes to zero or gives a finite number as $(s,\varepsilon)\to(0,0)$ depending on which limit is taken first.
We also observe that $s/(s+\varepsilon)^2$ eliminates the pole of $1/(s+\varepsilon)$ if $\varepsilon\to0$ is taken first. These are equivalent to saying that the hypersurface singularity of $s+\varepsilon=0$ is not removable and the limit of $(s,\varepsilon)\to(0,0)$ is not well-defined without specifying the order of the limits. This does not cause any problem because each regularization scheme has its own counter terms.

%%%%%%%%%%%%%%%%%%%%%%%%%%%%%%%%%%%%%%%%%%%%%%%%%%%%%%%%%%%%%%%%%%%%%%
\section{Prefactor for vacuum decay rate}\label{sec:vacuum_decay}
In this section, we demonstrate the calculation of functional determinants using examples of vacuum decay in quantum field theory. We consider a $D$-dimensional theory where there are two non-degenerated vacua, and calculate a decay rate of the higher (false, metastable) vacuum into the lower (true, stable) vacuum at zero temperature.

The decay rate of a metastable vacuum has been formulated in \cite{Coleman:1977py,Callan:1977pt}, and has the form of $\gamma=\mathcal Ae^{-\mathcal B}$,
where $\mathcal B$ is the Euclidean action of a bounce, and $\mathcal A$ is a prefactor having mass dimension $D$. Here, the bounce is a non-trivial $O(D)$ symmetric solution of the Euclidean equations of motion connecting the two vacua.

Meanwhile, the prefactor $\mathcal A$ represents the quantum corrections to the action, and has the form
\begin{equation}
    \mathcal A=J_{\rm trans}^D\ab|\frac{\det'\mathcal M}{\det\widehat{\mathcal M}}|^{-1/2},\label{eq:normal_prefactor}
\end{equation}
where $\mathcal M$ is the fluctuation operator around the bounce, and $\widehat{\mathcal M}$ is the one around the false vacuum. There appears a negative mode and thus we take the absolute value of the determinant. The prime indicates the subtraction of zero modes, which sets the dimension of $\mathcal A$. The zero modes are associated with the breaking of the translational invariance, and the corresponding Jacobian to the collective coordinates is given by $J_{\rm trans}^D=(\mathcal B/(2\pi))^{D/2}$. The details of the treatment of zero modes are in Appendix \ref{apx:zero}.

In the following, we examine two examples. The first one is scale invariant instantons in general dimensions, where we demonstrate that our method can indeed renormalize the functional determinant in higher dimensions. The other example is a generic renormalizable potential in four dimensions, where we compare the reference series that we have discussed so far.
%%%%%%%%%%%%%%%%%%%%%%%%%%%%%%%%%%%%%%%%%%%%%%%%%%%%%%%%%%%%%%%%%%%%%%
\subsection{Scale invariant potential}\label{subsec:scale_inv}
Let us consider an $O(N)$ symmetric scalar field theory in $D>2$ dimensions with a classically scale invariant potential,
\begin{equation}
    V(\phi)=\lambda (\phi\cdot\phi)^{\frac{D}{D-2}},\label{eq:scale_inv_potential}
\end{equation}
where $\phi$ is a vector of $N$ real scalars and $\lambda<0$ is a coupling constant. Notice that, although the potential is monotonic in $|\phi|$, the false vacuum at $|\phi|=0$ is stable against fluctuations due to the gradient term. Notice also that the potential of $|\phi|$ is analytic at $|\phi|=0$ for $D=3,4$ and $6$. Otherwise, the second derivative of the potential is continuous, which is actually enough for functional determinants to be well-defined.

Although Eq.~\eqref{eq:scale_inv_potential} involves multiple fields, the trajectory of the bounce is always straight in the field space in the case of scale invariant potentials \cite{Oda:2019njo}. Thus, the equations of motion are reduced to those of a single field $\bar h(r)=|\phi(r)|$. 
The bounce for the scale invariant potential is known as the Fubini-Lipatov instanton \cite{Fubini:1976jm}, which can be generalized to arbitrary $D>2$ dimensions as mentioned in \cite{Barbon:2010gn}. The bounce is given by
\begin{equation}
    \bar h(r)=\ab(\frac{2}{|\lambda|})^{\frac{D-2}{4}}\ab(\frac{D-2}{2}\frac{b}{r^2+b^2})^{\frac{D-2}{2}},
\end{equation}
which has the Euclidean action of
\begin{equation}
    \mathcal B=S_E[\bar h]=\ab(\frac{2|\lambda|}{\pi})^{1-\frac{D}{2}}\frac{(D-2)^{D-1}\pi\Gamma\ab(\frac{D}{2})}{\Gamma(D)}.
\end{equation}
Here, $b$ is an arbitrary positive constant determining the size of the bounce.

Since the bounce also breaks the scale invariance and the $O(N)$ symmetry, there appear additional zero modes, and all the configurations connected by these broken symmetries contribute to the decay rate.
Because of this, \eqref{eq:normal_prefactor} is slightly modified as
\begin{equation}
    \mathcal A=\mathcal V_{\rm group}\int_0^\infty\odif{b}J_{\rm trans}^DJ_{\rm scale}J_{\rm group}^{N-1}\ab|\frac{\det'\mathcal M^{(h)}}{\det\widehat{\mathcal M}^{(h)}}|^{-1/2}\ab(\frac{\det'\mathcal M^{({\rm NG})}}{\det\widehat{\mathcal M}^{({\rm NG})}})^{-(N-1)/2},\label{eq:prefactor_si}
\end{equation}
where $(h)$ indicates the Higgs modes toward the direction of $\bar h$, and $({\rm NG})$ indicates the Nambu-Goldstone (NG) modes. Here, $J_{\rm scale}$ and $J_{\rm group}$ are the Jacobians for the scale transformation and the broken internal symmetry, respectively, and $\mathcal V_{\rm group}=2\pi^{N/2}/\Gamma(N/2)$ is the group volume of the moduli space\footnote{The Jacobian of $J_{\rm group}$ takes care of a single flat direction, and we have $N-1$ flat directions. The broken group volume $\mathcal V_{\rm group}$ should be measured according to $J_{\rm group}$ following Appendix \ref{apx:zero}.}.
The integral over $b$ diverges at the one-loop level as we will see that the integrand is a power of $b$. However, higher-loop corrections may make it finite depending on the beta functions of the theory, and it indeed becomes finite in the standard model \cite{Andreassen:2017rzq,Chigusa:2017dux,Chigusa:2018uuj}. In this paper, we compute only the determinants and the Jacobians to avoid this additional complication.
%%%%%%%%%%%%%%%%%%%%%%%%%%%%%%%%%%%%%%%%%%%%%%%%%%%%%%%%%%%%%%%%%%%%%%
\subsubsection{Functional determinants of radial operators}
The fluctuation operators for the Higgs and NG modes are written in a uniform way as
\begin{align}
    \mathcal M^{(\kappa)}=-\partial^2+\kappa\ab(\frac{b}{r^2+b^2})^2,~\widehat{\mathcal M}=-\partial^2,
\end{align}
with $\kappa=-D(D+2)$ for the Higgs modes and $\kappa=-D(D-2)$ for the NG modes. Here, $\widehat{\mathcal M}$ is common to both. We also use the notation $\mathcal M^{(h)}$ and $\mathcal M^{({\rm NG})}$.

The solutions of Eq.~\eqref{eq:GY} are given by
\begin{align}
    \psi_\nu^{(\kappa)}(r)&=r^\nu\ab(\frac{b^2+r^2}{b^2})^{1+z(\kappa)}{}_2F_1\ab(1+z(\kappa),\nu+1+z(\kappa);\nu+1;-\frac{r^2}{b^2}),\\
    \hat \psi_\nu(r)&=r^\nu,
\end{align}
with
\begin{equation}
    z(\kappa)=-\frac12\ab(1-\sqrt{1-\kappa}).
\end{equation}
Here, ${}_2F_1(a,b;c;z)$ is the hypergeometric function.

Substituting these solutions into Eq.~\eqref{GY-theorem}, we obtain
\begin{equation}
    R_\nu(\kappa)=\frac{\Gamma\ab(\nu+1)\Gamma\ab(\nu)}{\Gamma\ab(\nu+1+z(\kappa))\Gamma\ab(\nu-z(\kappa))}.\label{eq:r_scale_inv}
\end{equation}

\subsubsection{Large angular momentum behavior}
A non-trivial check of our reference series is to see if its large $\nu$ behavior agrees with that of the determinant. The large $\nu$ expansion of Eq.~\eqref{eq:r_scale_inv} is given by
\begin{align}
    \ln R_\nu(\kappa)&=\frac{\kappa}{4\nu}-\frac{\kappa^2}{96\nu^3}+\frac{\kappa^2(\kappa+2)}{960\nu^5}-\frac{\kappa^2(3\kappa^2+16\kappa+32)}{21504\nu^7}+\frac{\kappa^2(\kappa+4)(\kappa^2+6\kappa+24)}{46080\nu^9}\nonumber\\
    &\hspace{3ex}-\frac{\kappa^2(\kappa^4+16\kappa^3+136\kappa^2+640\kappa+1280)}{270336\nu^{11}}+\order(\nu^{-13}).\label{eq:expansion_scale_invariant}
\end{align}
This is compared with Eq.~\eqref{eq:hke_z0} with $a_{\max}=6$ divided by $d_\nu$,
\begin{align}
    -\frac{\eta'^{\rm HKC}_\nu(0;0)}{d_\nu}&=-\sum_{a=1}^{6}\Xi_{a,\nu}(0)\int_0^\infty\odif{r}r^{2a-1}B_a^{z=0}[m^2](r).
\end{align}
Here, $m^2=\kappa\ab(\frac{b}{r^2+b^2})^2$. Notice that $B_a^{z=0}[\hat m^2]=0$ for $\hat m^2=0$. The $r$ integral can be executed explicitly as
\begin{align}
    \int_0^\infty\odif{r}rB_1^{z=0}[m^2](r)&=-\frac{\kappa}{2},\\
    \int_0^\infty\odif{r}r^3B_2^{z=0}[m^2](r)&=\frac{\kappa^2}{24},\\
    \int_0^\infty\odif{r}r^5B_3^{z=0}[m^2](r)&=-\frac{\kappa^2(\kappa+12)}{360},\\
    \int_0^\infty\odif{r}r^7B_4^{z=0}[m^2](r)&=\frac{\kappa^2(3\kappa^2+128\kappa+1152)}{20160},\\
    \int_0^\infty\odif{r}r^9B_5^{z=0}[m^2](r)&=-\frac{\kappa^2(\kappa^3+100\kappa^2+2880\kappa+23040)}{151200},\\
    \int_0^\infty\odif{r}r^{11}B_6^{z=0}[m^2](r)&=\frac{\kappa^2(\kappa^4+192\kappa^3+12192\kappa^2+294912\kappa+2211840)}{3991680}.
\end{align}
Using these, we reproduce all the coefficients of Eq.~\eqref{eq:expansion_scale_invariant} up to $\order(\nu^{-12})$.
Notice that the above discussion is independent of spacetime dimensions as we factored out $d_\nu$.
Since $d_\nu=\order(\nu^{D-2})$, we find $d_\nu\ln R_\nu(\kappa)+\eta'^{\rm HKC}_\nu(0;0)=\order(\nu^{D-15})$, which shows that our formula with $a_{\max}=6$ can indeed regularize functional determinants up to $D=13$ for arbitrary $\kappa$.

\subsubsection{Vacuum decay rate}
Finally, we present regularized determinants in various spacetime dimensions. From Eq.~\eqref{eq:prefactor_si}, the differential vacuum decay rate is given by
\begin{equation}
    \odv{\gamma}{b}=\frac{\mathcal V_{\rm group}}{b^{D+1}}\ab|\frac{1}{b^{2(D+1)}J_{\rm scale}^2J_{\rm trans}^{2D}}\frac{\det'\mathcal M^{({h})}}{\det\widehat{\mathcal M}}|^{-1/2}\ab(\frac{1}{J_{\rm group}^2}\frac{\det'\mathcal M^{({\rm NG})}}{\det\widehat{\mathcal M}})^{-(N-1)/2}\exp(-\mathcal B).
\end{equation}
Here, we factored out $1/b^{D+1}$ to make the rest dimensionless.

In the zeta function regularization, the above functional determinants are decomposed as
\begin{align}
    \ln\left|\frac{1}{b^{2(D+1)}J_{\rm scale}^2J_{\rm trans}^{2D}}\left.\frac{\det'\mathcal M^{({h})}}{\det\widehat{\mathcal M}}\right|_{\zeta}\right|&=\ab|\frac{1}{b^2J^2_{\rm scale}}\ln\frac{\det'\mathcal M_{D/2-1}^{(h)}}{\det\widehat{\mathcal M}_{D/2-1}}|+\eta'^{{\rm HKC} (h)}_{D/2-1}(0;0)\nonumber\\
    &\hspace{3ex}+D\ln\ab(\frac{1}{b^2J^2_{\rm trans}}\frac{\det'\mathcal M_{D/2}^{(h)}}{\det\widehat{\mathcal M}_{D/2}})+\eta'^{{\rm HKC} (h)}_{D/2}(0;0)\nonumber\\
    &\hspace{3ex}+\sum_{\nu=D/2+1}^\infty\ab[d_\nu\ln\ab(\frac{\det\mathcal M_{\nu}^{(h)}}{\det\widehat{\mathcal M}_{\nu}})+\eta'^{{\rm HKC} (h)}_{\nu}(0;0)],
\end{align}
and
\begin{align}
    \ln\ab(\frac{1}{J_{\rm group}^2}\left.\frac{\det'\mathcal M^{({\rm NG})}}{\det\widehat{\mathcal M}}\right|_{\zeta})&=\frac{1}{J^2_{\rm group}}\frac{\det'\mathcal M_{D/2-1}^{({\rm NG})}}{\det\widehat{\mathcal M}_{D/2-1}}+\eta'^{{\rm HKC} ({\rm NG})}_{D/2-1}(0;0)\nonumber\\
    &\hspace{3ex}+\sum_{\nu=D/2}^\infty\ab[d_\nu\ln\ab(\frac{\det\mathcal M_{\nu}^{({\rm NG})}}{\det\widehat{\mathcal M}_{\nu}})+\eta'^{{\rm HKC} ({\rm NG})}_{\nu}(0;0)].
\end{align}
Since these sums are absolutely convergent, one can compute it either numerically or analytically. Notice that we did not write the add-back term explicitly since ${S'}^{{\rm HKC}(h)}(0;0)={S'}^{{\rm HKC}({\rm NG})}(0;0)=0$ as we explained in Section \ref{sec:dimreg}.

The subtraction of zero modes requires careful treatment and is discussed in Appendix \ref{apx:zero}. The results are summarized as
\begin{align}
    \frac{1}{J^2_{\rm scale}}\frac{\det'\mathcal M_{D/2-1}^{(h)}}{\det\widehat{\mathcal M}_{D/2-1}}&=-\ab(\frac{2|\lambda|}{\pi})^{\frac{D}{2}-1}\frac{4b^2}{(D-2)^{D+1}}\Gamma\ab(\frac{D}{2}),\label{eq:zero_scale}\\
    \frac{1}{J^2_{\rm trans}}\frac{\det'\mathcal M_{D/2}^{(h)}}{\det\widehat{\mathcal M}_{D/2}}&=\ab(\frac{2|\lambda|}{\pi})^{\frac{D}{2}-1}\frac{b^2}{(D-2)^D}\Gamma\ab(\frac{D}{2}),\\
    \frac{1}{J^2_{\rm group}}\frac{\det'\mathcal M_{D/2-1}^{({\rm NG})}}{\det\widehat{\mathcal M}_{D/2-1}}&=\ab(\frac{2|\lambda|}{\pi})^{\frac{D}{2}-1}\frac{1}{(D-2)^{D-1}}\Gamma\ab(\frac{D}{2}).\label{eq:zero_group}
\end{align}

From the above formulas, we see that the determinants have the form of
\begin{align}
    \ln\left|\frac{1}{b^{2(D+1)}J_{\rm scale}^2J_{\rm trans}^{2D}}\left.\frac{\det'\mathcal M^{({h})}}{\det\widehat{\mathcal M}}\right|_{\zeta}\right|&=Q_1^{(h)}-Q_2^{(h)}\ln b^2+\frac12(D-2)(D+1)\ln|\lambda|,\\
    \ln\ab(\frac{1}{J_{\rm group}^2}\left.\frac{\det'\mathcal M^{({\rm NG})}}{\det\widehat{\mathcal M}}\right|_{\zeta})&=Q_1^{({\rm NG})}-Q_2^{({\rm NG})}\ln b^2+\frac12(D-2)\ln |\lambda|.
\end{align}
Here, $Q_1^{(h,{\rm NG})}$ and $Q_2^{(h,{\rm NG})}$ are constants given in Table \ref{tbl:numerical}. Notice that $Q_2^{(h,{\rm NG})}$ comes from the UV log divergence, but is embedded in an artificial IR log divergence of $\eta^{{\rm HKC}}_{\nu}(0,s)$ as explained in Section \ref{sec:full_hke}.
The results in the dimensional regularization can be obtained by the replacement $b^2\to b^2\mu^2 e^{\frac{1}{\bar\varepsilon}}$ in the right-hand side.

\begin{table}[t]
\centering
\begin{tabular}{c||c|c|c|c|c|c|c|c|c|c|c}
 $D$ & 3 & 4 & 5 & 6 & 7 & 8 & 9 & 10 & 11 & 12 & 13\\
  \hline\hline
 $Q_1^{(h)}$ & $12.4$ & $4.95$ & $-4.20$ & $-16.9$ & $-35.3$ & $-55.0$ & $-82.9$ & $-110$ & $-149$ & $-186$ & $-236$\\ 
 $Q_1^{({\rm NG})}$ & $0.973$ & $-0.547$ & $-1.19$ & $-2.93$ & $-4.29$ & $-6.11$ & $-7.80$ & $-9.69$ & $-11.6$ & $-13.6$ & $-15.6$ \\  
 $Q_2^{(h)}$ & $0$ & $3$ & $0$ & $\frac{18}{5}$ & $0$ & $\frac{790}{189}$ & $0$ & $\frac{265}{56}$ & $0$ & $\frac{23163}{4400}$ & $0$\\  
 $Q_2^{({\rm NG})}$& $0$ & $\frac13$ & $0$ & $\frac{3}{10}$ & $0$ & $\frac27$ & $0$ & $\frac{155}{567}$ & $0$ & $\frac{325}{1232}$ & $0$
\end{tabular}
\caption{Approximated values of $Q_1^{(h)}$ and $Q_1^{({\rm NG})}$ in three significant digits, and exact values of $Q_2^{(h)}$ and $Q_2^{({\rm NG})}$.}
\label{tbl:numerical}
\end{table}

The exact values of $Q_1^{(h,{\rm NG})}$ in three, four and six dimensions may be important since the potential becomes an integer power of $|\phi|$. Therefore, we compute them using the summation technique developed in \cite{Chigusa:2018uuj}, and summarize the results below.

For $D=3$,
\begin{align}
    Q_1^{(h)}&=\frac{71}{4}\ln2+\frac{7}{8\pi^2}\zeta(3),\\
    Q_1^{({\rm NG})}&=\frac{5}{4}\ln2+\frac{7}{8\pi^2}\zeta(3).
\end{align}

For $D=4$,
\begin{align}
    Q_1^{(h)}&=\frac32+12\ln A_G+\ln2+5\ln3-5\ln\pi,\\
    Q_1^{({\rm NG})}&=-\frac16+4\ln A_G-\frac13\ln2-\ln\pi,
\end{align}
where $A_G\simeq1.282$ is the Glaisher constant.
The results agree with \cite{Chigusa:2018uuj,Andreassen:2017rzq}.

For $D=6$,
\begin{align}
    Q_1^{(h)}&=\frac{439}{120}+22\ln A_G+7\ln15-\frac{209}{5}\ln2-14\ln\pi-4\zeta'(-3),\\
    Q_1^{({\rm NG})}&=\frac{19}{240}+5\ln A_G-\frac{22}{5}\ln2+\ln3-2\ln\pi-2\zeta'(-3).
\end{align}

\subsection{Generic potential}\label{subsec:generic_pot}
\begin{figure}[t]
    \centering
    \includegraphics[width=0.4\textwidth]{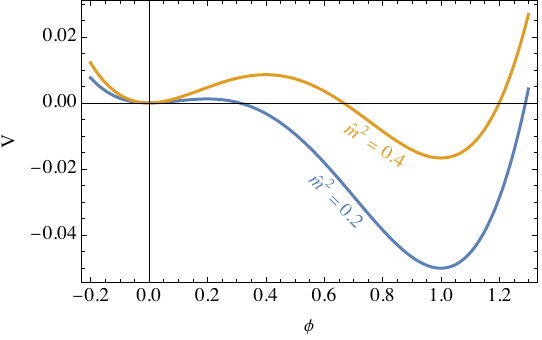}\hspace{1ex}
    \includegraphics[width=0.4\textwidth]{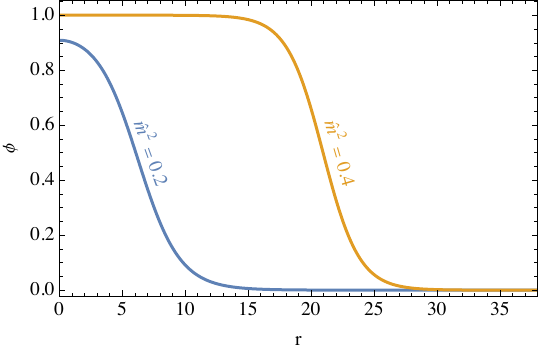}
    \caption{The potential (left) and the bounce (right) for $\hat m^2=0.2$ (blue) and $\hat m^2=0.4$ (orange) with $v=1$. The false vacuum is at $\phi=0$ and the true vacuum is at $\phi=1$.}
    \label{fig:bounce}
\end{figure}

Next, we consider a generic quartic potential in four dimensions:
\begin{equation}
    V(\phi)=\frac{1}{4}\phi^4-\frac{\hat m^2+v^2}{3v}\phi^3+\frac{\hat m^2}{2}\phi^2,
\end{equation}
where $0<\hat m^2<v^2/2$. The false vacuum is at $\phi=0$ and the true vacuum is at $\phi=v$. The two vacua are degenerate in the limit of $\hat m^2\to v^2/2$. See the left panel of Fig.~\ref{fig:bounce} for the shape of the potential. In this section, we consider $a_{\max}\leq5$ as the expression of $B_6^z[m^2]$ is too complicated and it is not worth computing $a=6$ in four dimensions.

The bounce is solved numerically by the overshoot-undershoot method \cite{Coleman:1977py} and is denoted by $\bar\phi$. The solution is shown in the right panel of Fig.~\ref{fig:bounce}.
Then, the fluctuation operators are given by $\mathcal M=-\partial^2+V''(\bar\phi)$, and $\widehat{\mathcal M}=-\partial^2+\hat m^2$.

In this subsection, we compare the $\nu$ dependence of $d_\nu\ln R_\nu$ after subtracting various reference series,
\begin{align}
    d_\nu\ln R_\nu^{\rm FD}|_{\rm fin}&=d_\nu\ln R_\nu-\sum_{a=1}^{a_{\rm max}}d_\nu[\ln R_\nu]_{\rm FD}^{(a)},\label{eq:fd_sub}\\
    d_\nu\ln R_\nu^{{\rm RWKB}}|_{\rm fin}&=d_\nu\ln R_\nu+\delta\Omega_\nu,\label{eq:rwkb_sub}\\
    d_\nu\ln R_\nu^{{\rm LAM}}|_{\rm fin}&=d_\nu\ln R_\nu+\eta'^{\rm LAM}_\nu(0),\\
    d_\nu\ln R_\nu^{{\rm HKC}}|_{\rm fin}&=d_\nu\ln R_\nu+\eta'^{\rm HKC}_\nu(z;0),
\end{align}
where $\ln R_{D/2}$ is understood as a value after subtracting the translational zero mode (see Appendix \ref{apx:zero}), and we take the absolute value of the determinants for $\nu=D/2-1$. The first and the second reference series are explained below.

The reference series in Eq.~\eqref{eq:fd_sub} is that of the Feynman diagrammatic approach \cite{Baacke:2003uw}. Here, $[\ln R_\nu]^{(a)}_{\rm FD}$ corresponds to the $a$ insertions of $\delta m^2$'s to a Feynman diagram, which can be calculated as
\begin{align}
    [\ln R_\nu]^{(1)}_{\rm FD}&=\frac{\psi_\nu^{(1)}(\infty)}{\hat\psi_\nu(\infty)},\\
    [\ln R_\nu]^{(2)}_{\rm FD}&=\frac{\psi_\nu^{(2)}(\infty)}{\hat\psi_\nu(\infty)}-\frac12\ab(\frac{\psi_\nu^{(1)}(\infty)}{\hat\psi_\nu(\infty)})^2,
\end{align}
where
\begin{align}
    \ab[\partial_r^2+\frac{1}{r}\partial_r-\frac{\nu^2}{r^2}-\hat m^2]\psi_\nu^{(a)}&=\delta m^2(r)\psi_\nu^{(a-1)},\label{eq:fd_ode}
\end{align}
with $\lim_{r\to0}\frac{\psi^{(a)}_\nu(r)}{r^{\nu}}=0$ for $a=1,2$, and $\psi_\nu^{(0)}=\hat\psi_\nu=I_\nu(\hat m r)$. Here, we only compute $a=1$ and $2$ for the Feynman diagrammatic approach because $a=3$ suffers from a difficulty in the computation of the corresponding add-back term.

The reference series in Eq.~\eqref{eq:rwkb_sub} is that of the radial WKB approach \cite{Hur:2008yg}. Since they only give the quantity after summing over $\nu$, we take a difference to get the contribution to each angular momentum, {\it i.e.}
\begin{equation}
    \delta \Omega_\nu=\int_0^\infty\odif{r}\ab[Q_{{\rm log},l}(r)-Q_{{\rm log},l-1}(r)+\sum_{n=-N_{\rm RWKB}}^{2}\ab[Q_{n,l}(r)l^n-Q_{n,l-1}(r)\ab(l-1)^n]]_{l=\nu-1},\label{eq:rwkb_sub_def}
\end{equation}
where $Q_{{\rm log},l}(r)$ and $Q_{n,l}(r)$ are functions of $l$ and $r$ defined in \cite{Hur:2008yg}. Here we ignore $l\leq1$ since the above formula becomes singular. Since $Q_{{\rm log},l}(r)$ depends on the renormalization scale, we take $\mu=1$ in our calculation. The order of the approximation is determined by $N_{\rm RWKB}$, and these functions have been calculated up to $N_{\rm RWKB}=4$. We treat $Q_{{\rm log},l}(r)$ as an $\order(\nu^0)$ contribution; we omit it from $\delta \Omega_\nu$ for $N_{\rm RWKB}<0$.
 
For convenience, we present the explicit formulas for our reference series at $a_{\max}=3$ for $D=4$:
\begin{align}
    -\eta'^{\rm LAM}_\nu(0)&=\frac{\nu}{2}\int_0^\infty\odif{r}r(m^2-\hat m^2)-\frac{1}{8\nu}\ab(1+\frac{1}{\nu^2})\int_0^\infty\odif{r}r^{3}(m^4-\hat m^4)\nonumber\\
    &\hspace{3ex}+\frac{1}{16\nu^3}\int_0^\infty\odif{r}r^{5}\ab(m^6+\frac12(\partial_rm^2)^2-\hat m^6),\label{eq:lam_a3}\\
    -\eta'^{\rm HKC}_\nu(z;0)&=\nu^2\int_0^\infty\odif{r}r\ab(m^2-\hat m^2)\mathcal I_{1,\nu}(z,r)\nonumber\\
    &\hspace{3ex}-\frac{\nu^2}{2}\int_0^\infty\odif{r}r\ab((m^2-z)^2-(\hat m^2-z)^2)\mathcal I_{2,\nu}(z,r)\nonumber\\
    &\hspace{3ex}+\frac{\nu^2}{6}\int_0^\infty\odif{r}r\ab((m^2-z)^3+\frac12(\partial_rm^2)^2-(\hat m^2-z)^3)\mathcal I_{3,\nu}(z,r).\label{eq:hkc_a3}
\end{align}

We will compare the reference series based on $C$\footnote{For brevity, we do not explicitly present the values of $C$ in the subsequent analysis, as the differences are readily discernible from the figures.} and $k$ of
\begin{equation}
    |d_\nu\ln R_\nu|_{\rm fin}|<C\nu^{-k},\label{eq:conv_par}
\end{equation}
for a sufficiently large $\nu$.
Since the computational cost is dominated by the evaluation of $\ln R_\nu$, a faster falloff (larger $k$) and a smaller error (smaller $C$) allow for an earlier truncation of the sum over $\nu$ without sacrificing accuracy, thereby improving the overall efficiency of the numerical computation.

\begin{figure}[t]
    \centering
    \includegraphics[width=0.4\textwidth]{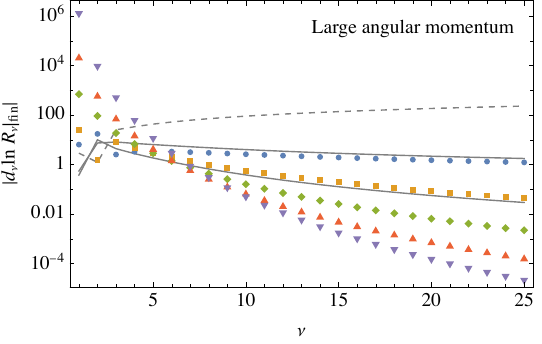}\hspace{1ex}
    \includegraphics[width=0.4\textwidth]{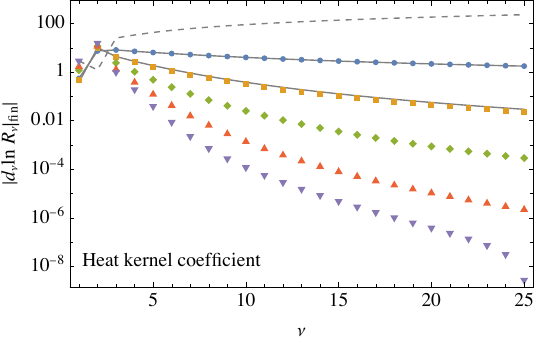}
    \caption{Absolute value of $d_\nu\ln R_\nu^{\rm fin}$ using the large angular momentum subtraction (left) and the heat kernel coefficient subtraction with $z=\hat m^2$ (right). The parameters are $\hat m^2=0.2$ and $v=1$, corresponding to a thick-wall bounce. Markers indicate $a_{\rm max}=1$ (blue circle), $2$ (orange square), $3$ (green diamond), $4$ (red triangle), and $5$ (purple inverted triangle). The dashed line shows the original $\ln R_\nu$ before subtraction. The two solid lines correspond to the Feynman diagrammatic subtraction for $a_{\max}=1$ and $2$ (from above).}
    \label{fig:numerical-thick}
\end{figure}

\begin{figure}[t]
    \centering
    \includegraphics[width=0.4\textwidth]{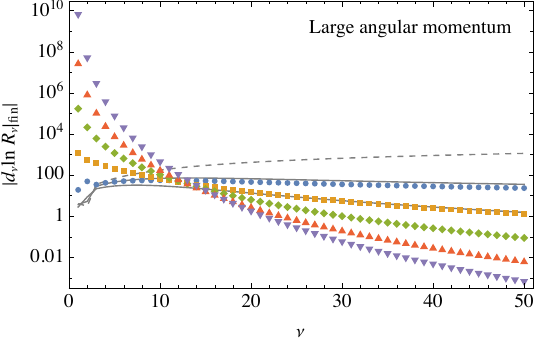}\hspace{1ex}
    \includegraphics[width=0.4\textwidth]{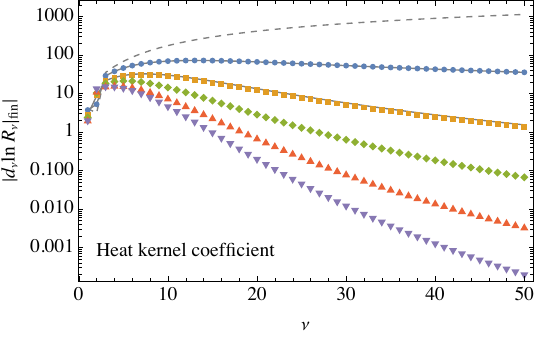}
    \caption{The same figure as in Fig.~\ref{fig:numerical-thick}, but for $\hat m^2=0.4$ and $v=1$, which corresponds to a thin-wall bounce.}
    \label{fig:numerical-thin}
\end{figure}

In the left panel of Figs.~\ref{fig:numerical-thick} and \ref{fig:numerical-thin}, we show $d_\nu\ln R_\nu^{\rm LAM}|_{\rm fin}$.
We also plot $d_\nu\ln R_\nu^{\rm FD}|_{\rm fin}$ with the gray solid lines for comparison. The blue circle points and the orange square points correspond to $a_{\max}=1$ and $2$, respectively, which are the same as the ones obtained in \cite{Dunne:2006ct}. The other points show $a_{\max}=3,4$ and $5$, which are our new results. As can be seen, the higher $a_{\max}$, the more suppression at a large $\nu$ as the exponent in Eq.~\eqref{eq:conv_par} is given by $k=2a_{\max}-1$.
This enables truncating the series at a smaller $\nu$, reducing the computational costs. A potential drawback would be that $\eta'^{\rm LAM}_\nu(0)$ at a small $\nu$ becomes gigantic, which may affect precision of the numerical computation. In practice, however, this issue can be easily solved by pre-including these small $\nu$ contributions in $S'^{\rm LAM}(0)$.

In the right panel of Figs.~\ref{fig:numerical-thick} and \ref{fig:numerical-thin}, we show $d_\nu\ln R_\nu^{\rm HKC}|_{\rm fin}$. Here we take $z=\hat m^2$, but it will turn out that this is not the best choice.
Even so, we observe that the large $\nu$ behavior is much better than that of $d_\nu\ln R_\nu^{\rm LAM}|_{\rm fin}$. This implies that our heat kernel coefficient approach gives a smaller $C$ in Eq.~\eqref{eq:conv_par} with the same exponent, $k=2a_{\max}-1$. The small $\nu$ behavior is also improved because the large $\nu$ expansion is not used in $\eta_\nu^{\rm HKC}(z;s)$. Notice that the blue points and the gray solid line agree because $\eta'^{\rm HKC}_\nu(\hat m^2;0)=-d_\nu[\ln R_\nu]_{\rm FD}^{(1)}$ for $a_{\max}=1$ and $z=\hat m^2$.

We highlight the significant reduction in computational cost achieved by our approach. In the Feynman diagrammatic method, it is necessary to solve additional differential equations, as given in Eq.~\eqref{eq:fd_ode}, for each value of $\nu$. In contrast, the computation of $d_\nu\ln R_\nu^{\rm LAM}|_{\rm fin}$ requires only a single integral, which can be reused for all $\nu$. For $d_\nu\ln R_\nu^{\rm HKC}|_{\rm fin}$, a one-dimensional integral must be performed for each $\nu$, but this remains considerably more efficient than solving differential equations.
Moreover, increasing $a_{\max}$ further reduces the number of subdeterminants that must be evaluated. For instance, to achieve an accuracy of $0.1$ in the thick-wall case, the Feynman diagrammatic approach requires computation up to $\nu \simeq 16$. By employing the $a_{\max}=3$ formulas, this requirement is reduced to $\nu\simeq11$ for $d_\nu\ln R_\nu^{\rm LAM}|_{\rm fin}$ and to $\nu\simeq6$ for $d_\nu\ln R_\nu^{\rm HKC}|_{\rm fin}$.
Even though the formulas for $a_{\max}=3$ remain very simple, as shown in Eqs.~\eqref{eq:lam_a3} and \eqref{eq:hkc_a3}, the resulting reduction in computational effort is substantial.

\begin{figure}[t]
    \centering
    \includegraphics[width=0.4\textwidth]{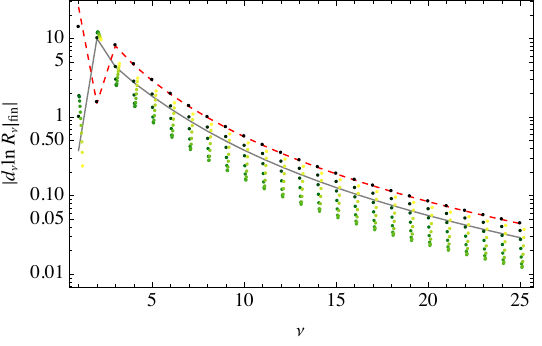}\hspace{1ex}
    \includegraphics[width=0.4\textwidth]{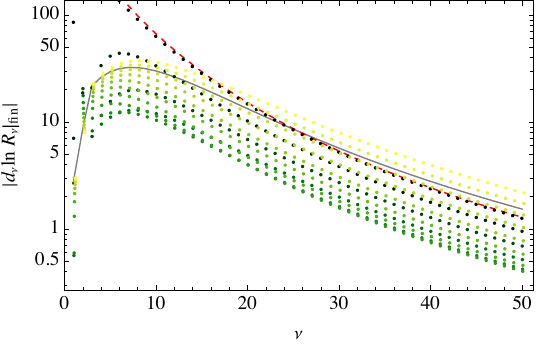}
    \caption{The $z$ dependence of $d_\nu\ln R_\nu^{\rm HKC}|_{\rm fin}$ for $a_{\max}=2$. We plotted $z=0.01\hat m^2$ and from $z=0.1\hat m^2$ (black) to $z=1.2\hat m^2$ (yellow) with step of $0.1\hat m^2$. The left panel is for $\hat m^2=0.2$, and the right panel is for $\hat m^2=0.4$. We shifted $\nu$ slightly for different $z$ to improve visibility. The gray solid curve corresponds to $\ln R_\nu^{\rm FD}|_{\rm fin}$ and the red dashed one to $\ln R_\nu^{\rm LAM}|_{\rm fin}$. The lowest points correspond to $z=0.12$ for $\hat m^2=0.2$ and $z=0.2$ for $\hat m^2=0.4$.}
    \label{fig:mass_dependence}
\end{figure}

In Fig.~\ref{fig:mass_dependence}, we show the $z$ dependence of $d_\nu\ln R_\nu^{\rm HKC}|_{\rm fin}$ for $a_{\max}=2$. We also plot $d_\nu\ln R_\nu^{\rm FD}|_{\rm fin}$ and $d_\nu\ln R_\nu^{\rm LAM}|_{\rm fin}$ for $a_{\max}=2$ with the black solid line and the red dashed line, respectively.
As $z\to0$, $d_\nu\ln R_\nu^{\rm HKC}|_{\rm fin}$ approaches $d_\nu\ln R_\nu^{\rm LAM}|_{\rm fin}$ since they only differ by the large $\nu$ expansion of Eq.~\eqref{eq:large_nu_exp} in this limit. We have omitted the $z=0$ data points, as they closely coincide with those at $z=0.01\hat m^2$, except for the lowest-angular-momentum contributions.
For $z=\hat m^2$, $d_\nu\ln R_\nu^{\rm HKC}|_{\rm fin}$ is close to $d_\nu\ln R_\nu^{\rm FD}|_{\rm fin}$ since they both use the same mass in the free Green's function for the expansion. Meanwhile, choosing a value of $z$ that is too large can deteriorate the large $\nu$ behavior, since the asymptotic regime of the modified Bessel functions is only reached for $\nu \gg z r_*$, where $r_*$ denotes a characteristic scale of the bounce. The optimal choice of $z$ seems to be intermediate in our examples and sits around $z\simeq 0.5\hat m^2$ for both thick-wall and thin-wall cases.

Let us discuss a systematic approach for selecting the optimal value of $z$ prior to numerical evaluation. This procedure is particularly advantageous for efficient and accurate numerical implementations.
We utilize the fact that $d_\nu\ln R_\nu^{\rm HKC}|_{\rm fin}$ includes a part of higher-order terms of $d_\nu\ln R_\nu^{\rm LAM}|_{\rm fin}$ as we can see from Eq.~\eqref{eq:small_mhat_exp}. We take $z$ so that
\begin{align}
    &\sum_{a=1}^{a_{\max}}\frac{(-z)^{a_{\max}+1-a}}{(a_{\max}+1-a)!}\int_0^\infty\odif{r}r^{2a_{\max}+1}\ab(B_a[m^2-z]-B_a[\hat m^2-z])\nonumber\\
    &\sim \int_0^\infty\odif{r}r^{2a_{\max}+1}\ab(B_{a_{\max}+1}[m^2]-B_{a_{\max}+1}[\hat m^2])
\end{align}
is satisfied as much as possible, as long as such $z$ is not too large. Since the difference of the right-hand side and the left-hand side is the net next order correction, this minimizes the next order correction and makes $C$ in Eq.~\eqref{eq:conv_par} smaller.
For $a_{\max}=2$, there is no exact solution, but we minimize their difference and obtain $z\simeq0.13$ for $\hat m^2=0.2$, and $z\simeq0.19$ for $\hat m^2=0.4$, which are in agreement with the lowest points in Fig.~\ref{fig:mass_dependence}. For $a_{\max}=3$, there exists a solution (around the best $z$ for $a_{\max}=2$), meaning that it completely eliminates the next order correction, {\it i.e.} $C=0$, and promotes the reference series equivalent to $a_{\max}=4$ without additional computational costs.

\begin{figure}[t]
    \centering
    \includegraphics[width=0.4\textwidth]{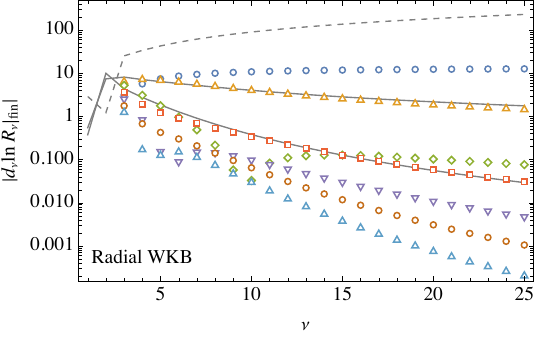}\hspace{1ex}
    \includegraphics[width=0.4\textwidth]{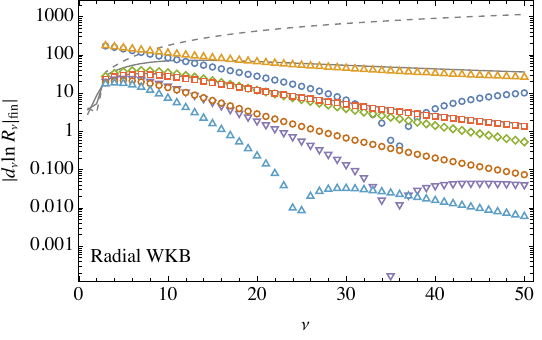} 
    \caption{Absolute value of $d_\nu\ln R_\nu^{\rm RWKB}|_{\rm fin}$ with the radial WKB approach. The left panel is for $\hat m^2=0.2$, and the right panel is for $\hat m^2=0.4$. Markers indicate $N_{\rm RWKB}=-2$ (blue circle), $-1$ (orange triangle), $0$ (green diamond), $1$ (red square), $2$ (purple inverted triangle), $3$ (brown circle), and $4$ (cyan triangle). The dashed line and the solid lines are the same as in Fig.~\ref{fig:numerical-thick}. The first few angular momenta are not shown since they are singular due to the definition of Eq.~\eqref{eq:rwkb_sub_def}.}
    \label{fig:rwkb}
\end{figure}

Finally, we discuss the reference series of the radial WKB approach \cite{Hur:2008yg}.
A correspondence between $N_{\rm RWKB}$ and $a_{\max}$ is given by $N_{\rm RWKB}\leftrightarrow2a_{\max}-3$, {\it e.g.} $N_{\rm RWKB}=4$ in their approach provides better exponent $k$ than $a_{\max}=3$ in our method, but is outperformed by $a_{\max}=4$. We note that the expressions required for $N_{\rm RWKB}=4$ already span approximately a page, whereas our formulation for $a_{\max}=4$ remains concise, requiring only about five lines. This highlights the relative simplicity and efficiency of our approach at higher orders. 
Fig.~\ref{fig:rwkb} shows the numerical results with the same parameters used in Figs.~\ref{fig:numerical-thick} and \ref{fig:numerical-thin}. The exponent in Eq.~\eqref{eq:conv_par} is given by $k=N_{\rm RWKB}+2$. Comparing with the other approaches for $N_{\rm RWKB}=3$ and $a_{\max}=3$, {\it i.e.} $k=5$, we observe that this approach gives almost the same performance as $d_\nu\ln R_\nu^{\rm LAM}|_{\rm fin}$, and is outperformed by $d_\nu\ln R_\nu^{\rm HKE}|_{\rm fin}$.

We also provide a Rust implementation of our method that can compute functional determinants for general potentials in four dimensions up to $a_{\max}=5$. The code is available at \cite{rust_github}.

%%%%%%%%%%%%%%%%%%%%%%%%%%%%%%%%%%%%%%%%%%%%%%%%%%%%%%%%%%%%%%%%%%%%%%
\section{Summary}\label{sec:summary}
In this paper, we first determined the higher-order coefficients in the large angular momentum expansion introduced in \cite{Dunne:2006ct}.
We found that these are related to the heat kernel coefficients with appropriate total derivatives, which we provide a systematic way to determine.
We explicitly give the reference series $\eta_\nu^{\rm LAM}(s)$ up to the sixth order of the heat kernel expansion, which suffices to regularize up to 13 dimensional functional determinants.

Next, we proposed a family of reference series $\eta_\nu^{\rm HKC}(z;s)$ that correspond to heat kernel coefficients. Since they are well-defined quantities beyond the angular momentum expansion, the computation of the add-back term becomes easier and the order counting becomes uniform in $\nu$.
We demonstrated the computation of functional determinants for the Fubini-Lipatov instantons in general dimensions and for a generic quartic potential in four dimensions. It is noteworthy that we successfully computed the functional determinants up to $D=13$, which has not been achieved before.
We found that our reference series better approximates the large angular momentum behavior of functional determinants than the other known approaches. We also found a way to choose $z$ that minimizes or even eliminates the next order coefficient.

The above two series are complementary.
With $\eta_\nu^{\rm HKC}(z\neq0;s)$, we need to execute one-dimensional integrals for each $\nu$, but we can truncate at a smaller $\nu$ in numerical computation. On the other hand, with $\eta_\nu^{\rm LAM}(s)$ or $\eta_\nu^{\rm HKC}(0;s)$, we only need to execute one dimensional integrals once, but we need to compute determinants for a larger $\nu$. For both reference series, the add-back terms $S^{\rm LAM}(s)$ and $S^{\rm HKC}(z;s)$ are much easier to calculate than in the Feynman diagrammatic approach, where we need to Fourier-transform $\delta m^2$ and convolute it with loop functions.
In particular, it is noteworthy that the reference series $\eta_\nu^{\rm HKC}(0;s)$ yields a vanishing add-back term in the zeta function regularization. This property simplifies the computation, as no additional formula for the add-back term is required.

Finally, we comment on some future directions.
One direction is to generalize our results to multiple fields, and to fermions and gauge bosons.
For $a_{\max}=2$, the generalization to multiple scalar fields is straightforward: one only needs to replace $m^2$ and $\hat m^2$ with corresponding mass matrices. 
For fermions and gauge bosons, such a straightforward\footnote{In obtaining the add-back term for gauge bosons, it should be noted that $\delta_\mu^\mu$ is also analytically continued in the dimensional regularization.} generalization is limited to $a_{\max}=2$ and $D\leq4$ due to a non-trivial subleading large $\nu$ behavior. 
Since $D=4$ results for fermions and gauge bosons using the large angular momentum subtraction are available in \cite{Baratella:2025dum}, it may be interesting to compare the results with our heat kernel coefficient approach.
The extension to larger $a_{\max}$ will be important for phenomenological applications and will be addressed in future work~\cite{YS2025}. Another promising direction is the generalization to more general spacetime metrics, which would be relevant for studying vacuum decay in the presence of black holes.

It would also be of mathematical interest to investigate whether a closed-form expression for $B_a^z[m^2]$ exists. Given that we have determined these coefficients up to $a=6$ and observed that they are considerably simpler than the original heat kernel coefficients, the existence of such a closed form appears plausible. Additionally, it may be worthwhile to clarify whether $\Delta_\nu(s)$ vanishes identically; although this does not impact the computation of functional determinants, it would provide further theoretical insight.

\begin{acknowledgments}
We would like to thank Takato Mori and Naritaka Oshita for useful discussions and comments.
Y.S. is supported by the Slovenian Research Agency under the research grant J1-4389. M.Y. is supported by IBS under the project code, IBS-R018-D3, and by JSPS Grant-in-Aid for Scientific Research Number JP23K20843.
\end{acknowledgments}

\appendix
\section{Matching at large angular momentum}
\label{apx:matching}
In this appendix, we determine the total derivative terms in $B_a^{z=0}[m^2]$ to match its large $\nu$ behavior to that of the functional determinant. We will do it in three steps: we first expand the functional determinant by mass insertions, then do the derivative expansion, and finally match the results to the heat kernel coefficients.
\subsection{Small mass expansion} 
Let us start from $R_\nu$ in Eq.~\eqref{GY-theorem} since this is how subdeterminants are computed practically.
We expand the solution $\psi_\nu(r)$ for a small $m^2$ as
\begin{align}
    \psi_\nu(r)&=r^\nu\ab[1+\psi^{(1)}_
    \nu(r)+\psi^{(2)}_
    \nu(r)+\cdots],
\end{align}
where $\psi_\nu^{(n)}(r)$ satisfies
\begin{equation}
    \ab[\partial_r^2+\frac{1+2\nu}{r}\partial_r]\psi_\nu^{(n+1)}= m^2\psi_\nu^{(n)},
\end{equation}
with $\psi_\nu^{(0)}(r)=1$ and
\begin{equation}
    \psi_\nu^{(n)}(0)=0,~\psi'^{(n)}_\nu(0)=0,
\end{equation}
for $n>0$.
By assigning a coupling $g$ to $m^2$, we can understand this as a coupling expansion and $\psi_\nu^{(n)}$ is $\order(g^n)$.

We put an artificial Dirichlet boundary at $|x|=r_\infty$, and examine
\begin{align}
    \tr P_\nu\ln(-\partial^2+m^2)-\tr P_\nu\ln(-\partial^2)&=\ln\frac{\det \ab(-\partial_r^2-\frac{1}{r}\partial_r+\frac{\nu^2}{r^2}+m^2)}{\det \ab(-\partial_r^2-\frac{1}{r}\partial_r+\frac{\nu^2}{r^2})}\nonumber\\
    &=\ln\frac{\psi_\nu(r_\infty)}{r_\infty^\nu}.
\end{align}
Expanding it for a small $m^2$, we obtain
\begin{equation}
    \tr P_\nu\ln(-\partial^2+m^2)-\tr P_\nu\ln(-\partial^2)\approx\sum_{n=1}^\infty[\ln \psi_\nu]^{(n)},\label{eq:partial_tr1}
\end{equation}
where $[\ln \psi_\nu]^{(n)}$ contains the $n$-th order terms.
For example,
\begin{align}
    [\ln \psi_\nu]^{(1)}&=\psi_\nu^{(1)}(r_\infty),\\
    [\ln \psi_\nu]^{(2)}&=\psi_\nu^{(2)}(r_\infty)-\frac12(\psi_\nu^{(1)}(r_\infty))^2,\\
    [\ln \psi_\nu]^{(3)}&=\psi_\nu^{(3)}(r_\infty)-\psi_\nu^{(1)}(r_\infty)\psi_\nu^{(2)}(r_\infty)+\frac13(\psi_\nu^{(1)}(r_\infty))^3.
\end{align}

On the other hand, we have a diagrammatic expansion by mass insertions,
\begin{align}
    \tr P_\nu\ln(-\partial^2+m^2)-\tr P_\nu\ln(-\partial^2)&\approx\sum_{n=1}^\infty\frac{(-1)^{n+1}}{n}\tr P_\nu\ab(\frac{1}{-\partial^2}m^2)^n.\label{eq:expansion_around_del2}
\end{align}
Here, the trace is evaluated in $|x|<r_\infty$, and we ignore boundary conditions since we are interested in $r_\infty\to\infty$.

Comparing Eqs.~\eqref{eq:partial_tr1} and \eqref{eq:expansion_around_del2}, we have\footnote{
For a small $n$, one can explicitly check this relation by using
\begin{align}
    \psi_\nu^{(n)}(r)&=\frac{1}{2\nu}\int_0^r\odif{r'}r'm^2(r')\psi_\nu^{(n-1)}(r')-\frac{1}{2\nu}\int_0^r\odif{r'}r'\ab(\frac{r'}{r})^{2\nu} m^2(r')\psi_\nu^{(n-1)}(r').
\end{align}
}
\begin{align}
    \lim_{r_\infty\to\infty}\frac{[\ln \psi_\nu]^{(n)}}{\frac{(-1)^{n+1}}{n}\tr\ab[P_\nu\ab(\frac{1}{-\partial^2}m^2)^n]}=1.
\end{align}
Thus, we can understand the large $\nu$ behavior of $\ln \psi_\nu(r_\infty)$ analyzing $\tr\ab[P_\nu\ab(\frac{1}{-\partial^2}m^2)^n]$. 
Since the UV divergence becomes weaker as $n$ increases, we only need to compute the first several terms.

\subsection{Derivative expansion}
In the following, we propose a derivative expansion of Eq.~\eqref{eq:expansion_around_del2}, which turns out to be a similar expansion as the heat kernel expansion.

Let us start with $n=1$. We insert the complete set of Eq.~\eqref{eq:complete_set_angular}, and obtain
\begin{align}
    \tr\ab[P_\nu\frac{1}{-\partial^2}m^2]&=\int_0^{r_\infty}\odif{r}rG_\nu(r,r)m^2(r)\nonumber\\
    &=\frac{1}{2\nu}\int_0^{r_\infty}\odif{r}rm^2(r).\label{eq:one_insertion}
\end{align}
Here, the radial Green's function is defined by
\begin{equation}
    G_\nu(r',r)=\int_0^\infty\odif{\lambda}\frac{J_\nu(\lambda r')J_\nu(\lambda r)}{\lambda}=
    \begin{cases}
        \frac{1}{2\nu}\ab(\frac{r}{r'})^\nu&r<r'\\
        \frac{1}{2\nu}\ab(\frac{r'}{r})^\nu&r\geq r'
    \end{cases}.
\end{equation}
Notice that the $\order(m^2)$ contribution of Eq.~\eqref{eq:one_insertion} is already of full order in the derivative expansion. Therefore the terms like $\partial^2 m^2,\partial^4m^2,\cdots$ never appear in $B_a^{z=0}[m^2]$.

For $n=2$, we compute
\begin{align}
    \tr\ab[P_\nu\ab(\frac{1}{-\partial^2}m^2)^2]&=\int_0^{r_\infty}\odif{r}r\int_0^{r_\infty}\odif{r_1}r_1G_\nu(r,r_1)m^2(r_1)G_\nu(r,r_1)m^2(r).
\end{align}
Since the integral is symmetric under $r\leftrightarrow r_1$, we only compute the integrals for $r>r_1$ and multiply the result by two. The equation becomes
\begin{align}
    \tr\ab[P_\nu\ab(\frac{1}{-\partial^2}m^2)^2]&=\frac{1}{2\nu^2}\int_0^{r_\infty}\odif{r}r^3m^2(r)\int_0^1\odif{x_1}x_1^{2\nu+1}m^2(rx),
\end{align}
where $x_1=r_1/r$.
For a large $\nu$, we see that $x_1^{2\nu+1}$ is almost zero except around $x_1=1$. This motivates us to execute the Taylor expansion around $x_1=1$. We obtain
\begin{align}
    \tr\ab[P_\nu\ab(\frac{1}{-\partial^2}m^2)^2]&\approx\left.\frac{1}{2\nu^2}\int_0^{r_\infty}\odif{r}r^3m^2(r)\sum_{q_1=0}^\infty \frac{1}{q_1!}\pdv[order=q_1]{}{\xi_1}m^2(r\xi_1)\int_0^1\odif{x_1}x_1^{2\nu+1}(x_1-1)^{q_1}\right|_{{\boldsymbol{\xi}}=1}\nonumber\\
    &=\left.\sum_{q_1=0}^\infty\frac{(-1)^{q_1}}{2\nu^2}\frac{\Gamma(2\nu+2)}{\Gamma(2\nu+q_1+3)}\int_0^{r_\infty}\odif{r}r^3\pdv[order=q_1]{}{\xi_1}m^2(r)m^2(r\xi_1)\right|_{{\boldsymbol{\xi}}=1}.
\end{align}
Here, $\boldsymbol{\xi}=1$ indicates taking all $\xi_i$ to $1$.
Now, we have a single integral over $r$, and the integrand is written only with $m^2$ and its derivatives. In addition, terms with more derivatives are suppressed by higher powers of $\nu$. These features smack of a relationship with the heat kernel expansion.

For $n=3$, we have three insertions, $m^2(r),m^2(r_1),m^2(r_2)$, and there are six ways to order $r,r_1$ and $r_2$. They all make the identical contribution, and we obtain
\begin{align}
    \tr\ab[P_\nu\ab(\frac{1}{-\partial^2}m^2)^3]&=\frac{3}{4\nu^3}\int_0^{r_\infty}\odif{r}r^5\int_0^1\odif{x_1}x_1^{2\nu+3}\int_0^{1}\odif{x_2}x_2^{2\nu+1}m^2(r)m^2(rx_1)m^2(rx_1x_2)\nonumber\\
    &\approx\sum_{q_1=0}^\infty\sum_{q_2=0}^\infty\frac{3}{4\nu^3}\frac{(-1)^{q_1+q_2}\Gamma(2\nu+2)\Gamma(2\nu+4)}{\Gamma(2\nu+q_2+3)\Gamma(2\nu+q_1+5)}\nonumber\\
    &\left.\hspace{6ex}\times\int_0^{r_\infty}\odif{r}r^5\pdv[order=q_1]{}{\xi_1}\pdv[order=q_2]{}{\xi_2}m^2(r)m^2(r\xi_1)m^2(r\xi_1\xi_2)\right|_{{\boldsymbol{\xi}}=1}.
\end{align}

For $n=4$, there are $24$ ways to order the positions of $m^2$'s. One can use symmetries to reduce the number of terms; (i) cyclic symmetry such as $r>r_1>r_2>r_3$ and $r_1>r_2>r_3>r$, (ii) flipping symmetry such as $r>r_1>r_2>r_3$ and $r<r_1<r_2<r_3$.
Using these, we effectively have three terms to be computed.
We obtain
\begin{align}
    \tr\ab[P_\nu\ab(\frac{1}{-\partial^2}m^2)^4]&=\frac{1}{\nu^4}\int_0^{r_\infty}\odif{r}r^7\int_0^1\odif{x_1}x_1^{2\nu+5}\int_0^1\odif{x_2}x_2^{2\nu+3}\ab(1+\frac{x_2^{2\nu}}{2})\int_0^{1}\odif{x_3}x_3^{2\nu+1}\nonumber\\
    &\hspace{6ex}\times m^2(r)m^2(rx_1)m^2(rx_1x_2)m^2(rx_1x_2x_3)\nonumber\\
    &\approx\sum_{q_1,q_2,q_3}\frac{1}{\nu^4}\frac{(-1)^{q_1+q_2+q_3}\Gamma(2\nu+2)\Gamma(2\nu+6)}{\Gamma(2\nu+q_3+3)\Gamma(2\nu+q_1+7)}\ab[\frac{\Gamma(2\nu+4)}{\Gamma(2\nu+q_2+5)}+\frac{\Gamma(4\nu+4)}{2\Gamma(4\nu+q_2+5)}]\nonumber\\
    &\left.\hspace{6ex}\times\int_0^{r_\infty}\odif{r}r^7\pdv[order=q_1]{}{\xi_1}\pdv[order=q_2]{}{\xi_2}\pdv[order=q_3]{}{\xi_3} m^2(r)m^2(r\xi_1)m^2(r\xi_1\xi_2)m^2(r\xi_1\xi_2\xi_3)\right|_{{\boldsymbol{\xi}}=1}.
\end{align}

For general $n$, we can obtain the formulas via
\begin{align}
    \tr\ab[P_\nu\ab(\frac{1}{-\partial^2}m^2)^n]&\approx\frac{n}{(2\nu)^n}\int_0^{r_\infty}\odif{r}r^{2n-1}\sum_{\sigma}\sum_{q_1,\cdots,q_{n-1}}\pdv[order=q_1]{}{\xi_1}\cdots\pdv[order=q_{n-1}]{}{\xi_{n-1}}\nonumber\\
    &\hspace{3ex}\times m^2(r)m^2(r\xi_1)m^2(r\xi_1\xi_2)\cdots m^2(r\xi_1\cdots\xi_{n-1})\nonumber\\
    &\hspace{3ex}\times \int_0^1\odif{x_1}x_1^{2n-3}\cdots\int_0^1\odif{x_{n-1}}x_{n-1}\nonumber\\
    &\hspace{3ex}\times \frac{(x_1-1)^{q_1}\cdots (x_{n-1}-1)^{q_{n-1}}}{q_1!\cdots q_{n-1}!} \frac{r_{\sigma_1}}{r}\frac{r_{\sigma_{n-1}}}{r}\nonumber\\
    &\hspace{3ex}\left.\times G(r_{\sigma_1},r_{\sigma_2})\cdots G(r_{\sigma_{n-2}},r_{\sigma_{n-1}})\right|_{\boldsymbol{\xi}=1}\nonumber\\
    &=\sum_{q=0}^\infty \mathcal W_{n,q}(\nu),
\end{align}
where $q=q_1+\cdots+q_{n-1}$, $\sigma$ is a permutation of $\{1,\cdots,n-1\}$ and $r_1=rx_1,r_2=rx_1x_2,r_3=rx_1x_2x_3,\cdots$.
Notice that $\mathcal W_{n,q}(\nu)=\order(\nu^{1-2n-q})$.

\subsection{Matching}
Now, we determine $B_a^{z=0}[m^2]$ such that $-\eta'^{\rm LAM}_\nu(0)/d_\nu$ has the same large $\nu$ behavior as Eq.~\eqref{eq:expansion_around_del2}. Let us define
\begin{align}
    \mathcal Y_{a_{\max}}(\nu)=-\sum_{a=1}^{a_{\max}}\Xi_{a,\nu}(s=0)\int_0^{r_\infty}\odif{r}r^{2a-1}B_a^{z=0}[m^2](r).
\end{align}
Since we know that $\Xi_{a,\nu}(0)$ is $\order(\nu^{1-2a})$ and that $\mathcal Y_{a_{\max}}(\nu)$ has maximally $a^{\max}$ insertions of $m^2$, we require
\begin{equation}
    \mathcal Y_{a_{\max}}(\nu)-\sum_{n=1}^{a_{\rm max}}\frac{(-1)^{n+1}}{n}\sum_{q=0}^{2(a_{\rm max}-n)}\mathcal W_{n,q}(\nu)=\order(\nu^{-2a_{\rm max}}).\label{eq:matching}
\end{equation}

We put an Ansatz of
\begin{equation}
    B_a^{z=0}[m^2]=b_a[m^2]+\mathcal D_a[m^2],
\end{equation}
where $b_a[m^2]$ is the original heat kernel coefficient, and $\mathcal D_a[m^2]$ includes total derivative terms written only with $m^2$ and its derivatives. We require that the coefficients of the total derivative terms do not depend on spacetime dimensions $D$.

Since the mass dimension of $\mathcal D_a[m^2]$ is $2a$, there are a finite number of possible total derivatives.
One can determine their coefficients substituting explicit functions for $m^2(r)$. For example, one can use $m^2(r)=1/(r^2+1)^p$ with $p\geq2$, and
\begin{equation}
    m^2(r)=\begin{cases}
        r^2(1-r)^p&(0<r<1)\\
        0&({\rm otherwise})
    \end{cases},
\end{equation}
with a large enough $p$ such that it becomes smooth enough.
After determining all the coefficients, we check that Eq.~\eqref{eq:matching} is satisfied for general $m^2(r)$.

For $B_1^{z=0}[m^2]$ to $B_5^{z=0}[m^2]$, we obtain Eqs.~\eqref{eq:large_b_from}-\eqref{eq:large_b_to}, which are unique. However, $B_6^{z=0}[m^2]$ is not unique because there are terms that do not contribute to $\mathcal Y_{a_{\max}}(\nu)$.
We give the explicit formulas of $B_6^{z=0}[m^2]$ and $\Delta B_6^z[m^2]$ below.
\begin{align}
    B_6^{z=0}[m^2]&=\frac{1}{720}\left[(m^2)^6+10(m^2)^3(\partial_\mu m^2)(\partial^\mu m^2)+\frac{31}{10}(\partial_\mu m^2)(\partial^\mu m^2)(\partial_\nu m^2)(\partial^\nu m^2)\right.\nonumber\\
    &\hspace{10ex}+\frac{62}{5}m^2(\partial_\mu\partial_\nu m^2)(\partial^\mu m^2)(\partial^\nu m^2)+\frac{18}{5}(m^2)^2(\partial_\mu\partial_\nu m^2)(\partial^\mu\partial^\nu m^2)\nonumber\\
    &\hspace{10ex}+\frac{3}{5}m^2(\partial_\mu m^2)(\partial^\mu m^2)(\partial^2m^2)+\frac{3}{5}(m^2)^2(\partial_\mu m^2)(\partial^\mu \partial^2m^2)\nonumber\\
    &\hspace{10ex}+\frac{3}{4}m^2(\partial_\mu\partial_\nu\partial_\rho m^2)(\partial^\mu\partial^\nu\partial^\rho m^2)+\frac{27}{70}m^2(\partial_\mu\partial_\nu \partial^2m^2)(\partial^\mu\partial^\nu m^2)\nonumber\\
    &\hspace{10ex}+\frac{9}{280}m^2(\partial_\mu \partial^2m^2)(\partial^\mu \partial^2m^2)+\frac{9}{280}m^2(\partial_\mu \partial^4m^2)(\partial^\mu m^2)\nonumber\\
    &\hspace{10ex}+\frac{8}{105}(\partial_\mu\partial_\nu\partial_\rho\partial_\sigma m^2)(\partial^\mu\partial^\nu\partial^\rho\partial^\sigma m^2)+\frac{23}{280}(\partial_\mu\partial_\nu\partial_\rho\partial^2 m^2)(\partial^\mu\partial^\nu\partial^\rho m^2)\nonumber\\
    &\hspace{10ex}+\frac{1}{56}(\partial_\mu\partial_\nu\partial^2 m^2)(\partial^\mu\partial^\nu\partial^2 m^2)+\frac{1}{56}(\partial_\mu\partial_\nu\partial^4 m^2)(\partial^\mu\partial^\nu m^2)\nonumber\\
    &\hspace{10ex}+\frac{1}{224}(\partial_\mu\partial^4 m^2)(\partial^\mu\partial^2 m^2)+\frac{1}{672}(\partial_\mu\partial^6 m^2)(\partial^\mu m^2)\nonumber\\
    &\hspace{10ex}+C_1(\partial^4 m^2)(\partial_\mu m^2)(\partial^\mu m^2)+C_2(\partial^2 m^2)(\partial_\mu \partial^2m^2)(\partial^\mu m^2)\nonumber\\
    &\hspace{10ex}+\ab(\frac{1747}{1260}-4C_1-4C_2)(\partial_\mu\partial_\nu m^2)(\partial^\mu\partial_\rho m^2)(\partial^\rho\partial^\nu m^2)\nonumber\\
    &\hspace{10ex}+\ab(\frac{8983}{2520}+2C_1+2C_2)(\partial_\mu\partial_\nu\partial_\rho m^2)(\partial^\mu m^2)(\partial^\rho\partial^\nu m^2)\nonumber\\
    &\hspace{10ex}+\ab(-\frac{9}{70}+4C_1+4C_2)(\partial^2 m^2)(\partial_\mu\partial_\nu m^2)(\partial^\mu\partial^\nu m^2)\nonumber\\
    &\hspace{10ex}+\ab(\frac{451}{504}-2C_1-3C_2)(\partial_\mu\partial^2 m^2)(\partial^\mu\partial_\nu m^2)(\partial^\nu m^2)\nonumber\\
    &\hspace{10ex}\left.+\ab(\frac{1769}{5040}-C_1)(\partial_\mu\partial_\nu\partial^2 m^2)(\partial^\mu m^2)(\partial^\nu m^2)\right].
\end{align}
Here, $C_1$ and $C_2$ are arbitrary constants. These ambiguities appear due to the following two vanishing linear combinations.
First, there is a combination that becomes zero when $m^2$ is a radial function,
\begin{align}
    0&=(\partial^4 m^2)(\partial_\mu m^2)(\partial^\mu m^2)-(\partial_\mu\partial_\nu\partial^2 m^2)(\partial^\mu m^2)(\partial^\nu m^2)\nonumber\\
    &\hspace{3ex}+(\partial_\mu\partial^2 m^2)(\partial^\mu\partial_\nu m^2)(\partial^\nu m^2)-(\partial^2 m^2)(\partial_\mu \partial^2m^2)(\partial^\mu m^2).
\end{align}
In addition, there is a combination that becomes zero after the $r$ integral when $m^2$ is a radial function,
\begin{align}
    &(\partial^4 m^2)(\partial_\mu m^2)(\partial^\mu m^2)-4(\partial_\mu\partial_\nu m^2)(\partial^\mu\partial_\rho m^2)(\partial^\rho\partial^\nu m^2)\nonumber\\
    &\hspace{3ex}+2(\partial_\mu\partial_\nu\partial_\rho m^2)(\partial^\mu m^2)(\partial^\rho\partial^\nu m^2)+4(\partial^2 m^2)(\partial_\mu\partial_\nu m^2)(\partial^\mu\partial^\nu m^2)\nonumber\\
    &\hspace{3ex}-2(\partial_\mu\partial^2 m^2)(\partial^\mu\partial_\nu m^2)(\partial^\nu m^2)-(\partial_\mu\partial_\nu\partial^2 m^2)(\partial^\mu m^2)(\partial^\nu m^2)\nonumber\\
    &=\frac{1}{r^{11}}\partial_r\ab[\frac{D-1}{3}r^9(\partial_rm^2)^2[3r(\partial_r^2m^2)+(D-3)(\partial_r m^2)]].
\end{align}

Once we obtain $B_6^{z=0}[m^2]$, we can compute $\Delta B_6^z[m^2]$ following Section \ref{sec:full_hke}.
The additional total derivative terms are given by
\begin{align}
    \Delta B_6^z[m^2]&=\frac{1}{720}\left[-\frac{z^2}{10}(\partial_\mu\partial^2m^2)(\partial^\mu m^2)-\frac{z^2}{10}(\partial_\mu\partial_\nu m^2)(\partial^\mu\partial^\nu m^2)\right.\nonumber\\
    &\hspace{10ex}-\frac{3z}{140}(\partial_\mu\partial^4m^2)(\partial^\mu m^2)-\frac{3z}{20}(\partial_\mu\partial_\nu\partial^2m^2)(\partial^\mu\partial^\nu m^2)\nonumber\\
    &\hspace{10ex}-\frac{3z}{140}(\partial_\mu\partial^2m^2)(\partial^\mu\partial^2 m^2)-\frac{3z}{28}(\partial_\mu\partial_\nu\partial_\rho m^2)(\partial^\mu\partial^\nu\partial^\rho m^2)\nonumber\\
    &\hspace{10ex}-\frac{3z}{10}m^2(\partial_\mu\partial^2m^2)(\partial^\mu m^2)-\frac{3z}{10}m^2(\partial_\mu\partial_\nu m^2)(\partial^\mu\partial^\nu m^2)\nonumber\\
    &\left.\hspace{10ex}-\frac{3z}{5}(\partial_\mu\partial_\nu m^2)(\partial^\mu m^2)(\partial^\nu m^2)-\frac{3z}{20}(\partial^2 m^2)(\partial_\mu m^2)(\partial^\mu m^2)\right].
\end{align}

%%%%%%%%%%%%%%%%%%%%%%%%%%%%%%%%%%%%%%%%%%%%%%%%%%%%%%%%%%%%%%%%%%%%%%
\section{Zero modes}\label{apx:zero}
The functional determinant around the bounce has zero modes because the bounce breaks symmetries.
For the Fubini-Lipatov instantons, we can explicitly see that some of subdeterminants become zero.
For the Higgs modes, the ratio of the determinants is given by
\begin{equation}
    \frac{\det\mathcal M_{l+D/2-1}^{(h)}}{\det\widehat{\mathcal M}_{l+D/2-1}}=\frac{\Gamma\ab(\frac{D}{2}+l)\Gamma\ab(\frac{D}{2}+l-1)}{\Gamma\ab(D+l)\Gamma\ab(l-1)}.
\end{equation}
As we can see, in $l=0$ and $1$, the functional determinant has zero modes since the bounce breaks the scale invariance and the translational invariance, respectively.
Similarly, for the NG modes, we have
\begin{equation}
    \frac{\det\mathcal M_{l+D/2-1}^{({\rm NG})}}{\det\widehat{\mathcal M}_{l+D/2-1}}=\frac{\Gamma\ab(\frac{D}{2}+l)\Gamma\ab(\frac{D}{2}+l-1)}{\Gamma\ab(D+l-1)\Gamma\ab(l)},
\end{equation}
which has a zero mode in $l=0$. Since there are $N-1$ NG fluctuation operators, we have $N-1$ zero modes. These zero modes arise from the breaking of $O(N)$ symmetry into $O(N-1)$ symmetry, which gives $N-1$ broken generators.

In the following, we explain the subtraction of zero modes in two cases; normalizable zero modes and non-normalizable zero modes.

\subsection{Normalizable zero mode}
In most cases, zero modes are normalizable and can be subtracted by adding a mass term and divided by the mass:
\begin{equation}
    \frac{\det'\mathcal M_\nu}{\det\widehat{\mathcal M}_\nu}=\lim_{\tilde\mu^2\to0}\frac{1}{\tilde\mu^2}\frac{\det\mathcal M_\nu+\tilde\mu^2}{\det\widehat{\mathcal M}_\nu}=\lim_{\tilde\mu^2\to0}\frac{1}{\tilde\mu^2}\frac{(0+\tilde\mu^2)(\omega_2+\tilde\mu^2)(\omega_3+\tilde\mu^2)\cdots}{\hat \omega_1\hat \omega_2\hat \omega_3\cdots},
\end{equation}
where the $\omega_i$ and $\hat\omega_i$ are the eigenvalues of $\mathcal M_\nu$ and $\widehat{\mathcal M}_\nu$, respectively.

Now, we make the connection to the Gelfand-Yaglom theorem. Recall that
\begin{equation}
    \frac{\det\mathcal M_\nu+\tilde\mu^2}{\det\widehat{\mathcal M}_\nu}=\frac{\Psi_\nu(\tilde\mu^2;r_\infty)}{\hat\psi_\nu(r_\infty)},
\end{equation}
with
\begin{equation}
    (\mathcal M_\nu+\tilde\mu^2)\Psi_\nu(\tilde\mu^2;r)=0.
\end{equation}
We can expand the solution as
\begin{equation}
    \Psi_\nu(\tilde\mu^2;r)=\psi_\nu(r)+\tilde\mu^2\check\psi_\nu(r)+\cdots,
\end{equation}
where $\check\psi_\nu$ satisfies
\begin{equation}
    \mathcal M_\nu\check\psi_\nu(r)=-\psi_\nu(r),
\end{equation}
with initial conditions $\lim_{r\to0}\check\psi_\nu(r)/r^\nu=0$. Here $\psi_\nu(r)$ is the original solution of \eqref{eq:GY}.

Taking $\tilde\mu^2\to0$, we obtain the determinant without the zero mode as
\begin{equation}
    \frac{\det'\mathcal M_\nu}{\det\widehat{\mathcal M}_\nu}=\frac{\check\psi_\nu(\infty)}{\hat\psi_\nu(\infty)}.\label{eq:det_prime}
\end{equation}
Notice that $\psi_\nu(\infty)/\hat\psi_\nu(\infty)=0$ due to the zero mode.
This method is used for the generic quartic potential in Subsection \ref{subsec:generic_pot}.

\subsection{Non-normalizable zero mode}
There are some cases where zero modes are non-normalizable and Eq.~\eqref{eq:det_prime} diverges. The zero modes for the scale invariance and the internal symmetry in the case of the Fubini-Lipatov instanton are such examples. Nevertheless, the divergence is superficial and is canceled by the Jacobian.
In the following, we explain how to directly compute the Jacobian and Eq.~\eqref{eq:det_prime} simultaneously following \cite{Chigusa:2018uuj}.

Suppose there is a symmetry
\begin{equation}
    \phi(x)\to\phi(x)+\theta G(x)+\order(\theta^2).
\end{equation}
Here, $\theta$ is the collective coordinate, with which the volume of the moduli space should be measured.
The zero mode is the fluctuation proportional to $G(x)$, {\it i.e.} $G(x)$ satisfies $\mathcal M_\nu G(x)=0$. Here, $\mathcal M_\nu$ is a matrix fluctuation operator acting on a vector $\phi-\bar\phi$, with $\bar\phi$ being the bounce.
The corresponding Jacobian from the path integral variables to the collective coordinates is given by
\begin{equation}
    J=\sqrt{\frac{\braket<G|G>}{2\pi}},
\end{equation}
where the inner product is defined as
\begin{equation}
    \braket<f|g>=\sum_{i=1}^N\int\odif[order=D]{x} f^*_i(x)g_i(x).
\end{equation}
This is easily derived from the path integral measure.
As we can see, the Jacobian diverges if the zero mode is non-normalizable.

Let us first introduce a function $\rho(r)$ such that
\begin{equation}
    \bra<G|\rho\ket|G>=2\pi.\label{eq:normalization}
\end{equation}
This becomes the substitute of $\braket<G|G>/J^2=2\pi$ in the case of normalizable zero modes.
Although $\braket<G|G>$ and $J$ diverges for a non-normalizable zero mode, this combination is always finite.

Using $\rho(x)$, we can calculate both of the Jacobian and the determinant prime simultaneously as
\begin{equation}
    \frac{1}{J^2}\frac{\det'\mathcal M_\nu}{\det\widehat{\mathcal M}_\nu}=\lim_{\tilde\mu^2\to0}\frac{1}{\tilde\mu^2}\frac{\det\mathcal M_\nu+\tilde\mu^2 \rho}{\det\widehat{\mathcal M}_\nu}=\lim_{\tilde\mu^2\to0}\frac{1}{\tilde\mu^2}\frac{(0+\tilde\mu^2/J^2)(\omega_2+\order(\tilde\mu^2))(\omega_3+\order(\tilde\mu^2))\cdots}{\hat \omega_1\hat \omega_2\hat \omega_3\cdots}.
\end{equation}

As in the previous subsection, we utilize the Gelfand-Yaglom theorem and obtain
\begin{equation}
    \frac{1}{J^2}\frac{\det'\mathcal M_\nu}{\det\widehat{\mathcal M}_\nu}=\frac{\check\psi^J_\nu(\infty)}{\hat\psi_\nu(\infty)},
\end{equation}
where
\begin{equation}
    \mathcal M_\nu\check\psi^J_\nu(r)=-\rho(r)\psi_\nu(r),\label{eq:psi_check_j}
\end{equation}
with initial conditions $\lim_{r\to0}\check\psi^J_\nu(r)/r^\nu=0$.

Notice that, if $\rho(r)$ is a constant and the zero mode is normalizable, Eq.\eqref{eq:normalization} sets $\rho(r)=1/J^2$, and we have $\check\psi^J_\nu(r)=\check\psi_\nu(r)/J^2$.

\subsubsection{Scale invariant instanton}
For scale invariant instantons in Subsection \ref{subsec:scale_inv}, it is convenient to take
\begin{equation}
    \rho(r)=c\ab(\frac{b}{r^2+b^2})^2,
\end{equation}
where $c$ is a constant to satisfy Eq.~\eqref{eq:normalization}.
Then, we do not need to solve Eq.~\eqref{eq:psi_check_j} since the addition of $\tilde\mu^2\rho$ corresponds to a shift of $\kappa$. We obtain
\begin{align}
    \frac{1}{J^2}\frac{\det'\mathcal M_\nu^{(\kappa_*)}}{\det\widehat{\mathcal M}_\nu}&=\lim_{\kappa\to\kappa_*}c\odv{}{\kappa}R_\nu^{(\kappa)},
\end{align}
where $\kappa_*=-D(D+2)$ for the Higgs modes and $\kappa_*=-D(D-2)$ for the NG modes. The coefficient is determined as
\begin{align}
    c_{\rm scale}=c_{\rm trans}&=\ab(\frac{8|\lambda|}{\pi})^{\frac{D}{2}-1}\frac{32b^2}{(D-2)^D\sqrt{\pi}}\Gamma\ab(\frac{D+3}{2}),\\
    c_{\rm group}&=\ab(\frac{8|\lambda|}{\pi})^{\frac{D}{2}-1}\frac{4}{(D-2)^{D-2}\sqrt{\pi}}\Gamma\ab(\frac{D+1}{2}).
\end{align}
Using these and Eq.~\eqref{eq:r_scale_inv}, we obtain Eqs.~\eqref{eq:zero_scale}-\eqref{eq:zero_group}.
%%%%%%%%%%%%%%%%%%%%%%%%%
\bibliographystyle{apsrev4-1}
\bibliography{det}
\end{document}